# A Physics-Informed Data-Driven Discovery for Constitutive Modeling of Compressible, Nonlinear, History-Dependent Soft Materials under Multiaxial Cyclic Loading

Alireza Ostadrahimi, Amir Teimouri, Kshitiz Upadhyay, Guoqiang Li[1]

Department of Mechanical and Industrial Engineering, Louisiana State University, Baton Rouge, LA

**Abstract**

We propose a general hybrid physics-informed machine learning framework for modeling nonlinear, history-dependent viscoelastic behavior under multiaxial cyclic loading. The approach is built on a generalized internal state variable-based visco-hyperelastic constitutive formulation, where stress is decomposed into volumetric, isochoric hyperelastic, and isochoric viscoelastic components. Gaussian Process Regression (GPR) models the equilibrium response, while Recurrent Neural Networks (RNNs) with Long Short-Term Memory (LSTM) units capture time-dependent viscoelastic effects. Physical constraints, including objectivity, material symmetry, and thermodynamic consistency, are enforced to ensure physically valid predictions. After developing the general form of the surrogate model based on tensor integrity bases and response functions, we employed the nonlinear Holzapfel differential viscoelastic model to generate training data. Two datasets, one for short-term and another for long-term relaxation, are constructed to span a wide range of material memory characteristics. The model is trained and tested under diverse multiaxial loading conditions, including different stretch levels applied independently in the longitudinal and transverse directions, varying strain rates, and both tension and compression states, even beyond the training domain. Energy dissipation is explicitly analyzed at different strain rates for both datasets to verify thermodynamic consistency through the second law. The results show that the proposed framework accurately captures complex, nonlinear, and rate-dependent material responses. Moreover, it demonstrates strong robustness to synthetic noise, enabling generalizable and physically consistent predictions under realistic and variable loading scenarios.

## 1 Introduction

Soft materials such as elastomers, biological tissues, and polymeric foams exhibit complex mechanical behavior characterized by large deformations, stress-strain nonlinearity, and history-dependent viscoelastic effects ([1], [2], [3], [4]). Understanding and accurately modeling this behavior is crucial for applications in biomedical engineering, soft robotics, and flexible electronics ([5], [6], [7], [8], [9], [10], [11], [12]). Traditional constitutive models, including linear viscoelasticity, finite linear viscoelasticity ([13],[14]), and quasi-linear viscoelasticity ([15], [16]) have served as fundamental tools for describing the mechanical response of such materials. These models often rely on linear superposition principles and Prony series representations to capture time-dependent stress relaxation and creep behavior. While effective in certain scenarios, they struggle to describe nonlinear viscoelasticity and finite deformation behavior typical of soft materials under large strains ([17], [18], [19]).

---

[1] lguoqi1@lsu.edu

To address these shortcomings, several finite nonlinear viscoelasticity theories have been developed. Most of these theories assume that the time-dependent response of materials is related to irreversible thermodynamic processes resulting from the evolution of the material's internal microstructure[20], [21], [22], [23], [24], [25]. Consequently, internal or external thermodynamic state variables are employed to account for rate/time-dependent mechanical responses [26], [27], [28], [29]. External state variable-based models employ objective strain rate tensors as thermodynamic state variables to efficiently describe strain rate stiffening or softening in material responses through a viscous dissipation potential [30], [31]. However, they assume viscous material behavior is dependent only on the instantaneous deformation rates (i.e., very recent history), and dependence on the entire previous loading history is neglected [32], [33]. Consequently, these models specialize in describing short-term responses (e.g., high-strain-rate deformations) [34], [35]. On the other hand, the internal state variable-based models describe the non-equilibrium part of the Helmholtz free energy function via a set of internal (i.e., hidden) variables that evolve in time during deformation, leading to time-dependent relaxation processes that are functions of the entire loading history ([1], [5], [36]). These models offer improved accuracy in capturing long-term (e.g., creep and relaxation) responses and have also been used in high-strain-rate loading applications under large deformations. Despite their promising aspects, calibration of internal state variable-based models requires extensive experimental data and expert knowledge, and their computational cost can become prohibitive in large-scale finite element simulations ([37], [38], [39]).

Accurate constitutive modeling is crucial for reliably predicting material behavior in finite element simulations, particularly for nonlinear and history-dependent materials. Traditional constitutive models rely on predefined mathematical formulations that require extensive calibration and iterative numerical integration ([40], [41]). While computational implementations (VUMAT/UMAT) exist, these models struggle with computational efficiency and prediction of complex responses ([42], [43]). Machine Learning (ML) eliminates the need for rigid strain-energy functions or viscoelastic kernels, learning complex material behavior directly from the experimental data for improved adaptability and automation ([44]). Traditional constitutive models also struggle with multi-physics effects, requiring complicated extensions for thermo-mechanical interactions and damage evolution. ML frameworks naturally learn these dependencies from multi-scale data, improving predictive efficiency ([45], [46]). Despite their promise, purely data-driven models present challenges, especially when experimental data are sparse or when the model is required to extrapolate to unobserved deformation paths ([47], [48], [49], [50], [51]). Without physical constraints, these models can produce non-physical stress predictions that violate fundamental principles such as objectivity and thermodynamic consistency ([52], [53]).

ML has emerged as a powerful tool for constitutive modeling ([51], [54], [55], [56]). Physics-constrained ML methods such as thermodynamics-based networks ([57], [58]), Bayesian constitutive models ([59]), and convex neural networks improve accuracy and physical consistency ([48], [60], [61], [62], [63]). Mechanics-informed models have addressed nonlinear viscoelasticity ([64], [65]), while hyperspace networks ([66]) and benchmarked methods ([67]) have shown enhanced robustness under limited data. Applications in biomedicine include risk prediction ([68]), disease progression ([69]), and skin growth modeling ([70]). ML has also been applied to plasticity without stress data ([71], [72], [73]) and fracture modeling ([46]), with hybrid approaches improving interpretability ([74], [75], [76]). Constitutive Artificial Neural Networks (CANNs) have been proposed to embed viscoelasticity into neural architectures ([55], [77], [78], [79], [80]), while Physics-Informed Neural Networks (PINNs) and Pretrained Audio Neural Networks (PANNs) have been used to enforce governing equations in problems like electro-elasticity and inelasticity ([62], [81], [82], [83]). Moreover, Bayesian Neural Networks (BNNs) provide uncertainty quantification under data scarcity ([72], [84], [85], [86], [87], [88], [89], [90], [91]). Other advances include Finite Element Artificial Neural Networks (FEANNs) ([92], [93], [94]), plasticity ([95]), interface mechanics ([96]), and hyperelasticity in metamaterials ([97], [98], [99]) and biomechanical systems ([48], [73]). Further, model-free methods have been proposed to address inelasticity ([83], [100], [101]), while sparse regression and iCNNs have been shown to improve parameter discovery ([63], [102]). Tools like EUCLID ([73]), Keras, and TensorFlow ([103]) support implementation of deep learning models by providing efficient computation, and built-in modules for designing, training, and evaluating neural networks. Recent extensions of FEANN ([93], [94]), POD-TANN ([104]), Bayesian CANNs and iCANNs ([80], [88]), enhance stability and efficiency but often lack sequential memory critical to viscoelastic modeling ([93]).

Gaussian Process Regression (GPR) is a supervised learning technique that formulates prediction as a problem of Bayesian inference over a distribution of functions ([105], [106]). It is a non-parametric model that is particularly well-suited for small datasets due to its ability to provide exact interpolation of training data and quantify uncertainty in predictions ([107]). In GPR, the similarity between input data points is defined through a kernel or covariance function, typically based on the Euclidean distance, which governs the smoothness and flexibility of the model output. In recent years, physics-informed GPR has gained significant attention as a powerful extension of this method, where physical laws, such as objectivity, isotropy, and thermodynamic consistency, are embedded directly into the learning framework. These models produce interpretable and physically consistent predictions and have shown strong performance in modeling complex, rate-independent material behavior ([107], [108], [109]). However, most existing physics-informed GPR approaches are limited to memoryless constitutive models and are not equipped to intrinsically handle history-dependent responses such as those seen in nonlinear viscoelasticity [110]. Addressing this gap, the present work employs a framework to effectively model nonlinear, rate-dependent, and history-sensitive responses in soft materials across a wide range of loading conditions while maintaining physical fidelity and generalizability.

A key breakthrough in constitutive modeling is the use of Recurrent Neural Networks (RNNs), particularly those incorporating Long Short-Term Memory (LSTM) units, which have demonstrated notable success in modeling history-dependent materials ([37], [111], [112]). Unlike traditional neural networks, RNNs inherently capture sequential dependencies, making them highly effective for predicting viscoelastic behavior under complex loading conditions, where stress evolution depends on the entire deformation history ([113], [114], [115]). Their ability to represent time-dependent responses and fading memory effects makes them particularly well-suited for learning viscoelastic constitutive laws ([116], [117]). These advancements have led to improved material descriptions for elastomers ([118], [119]) and temperature- and rate-dependent stress-strain models ([119]).

This study presents a physics-informed, data-driven framework that integrates GPR and RNN to model the nonlinear, history-dependent behavior of compressible soft materials under multiaxial cyclic loading. The framework is built upon a generalized internal state variable-based nonlinear visco-hyperelastic constitutive formulation. This allows for the efficient capture of large deformation responses across multiple timescales. To ensure physical reliability, the framework incorporates key physics-based constraints, including thermodynamic consistency, by enforcing the Clausius-Duhem inequality for isothermal processes. The model is rigorously validated for both its fitting and predictive capabilities across a wide range of loading scenarios, including rate-dependent responses, large deformations beyond the training domain, across tension and compression. Throughout, the framework consistently delivers high predictive accuracy under complex multiaxial loading conditions. Section 2 presents the generalized internal state variable-based visco-hyperelastic framework that captures both equilibrium and non-equilibrium stress components, providing a foundation for modeling complex material responses. In Section 3, we introduce a physics-informed, data-driven constitutive model that combines GPR for the equilibrium hyperelastic behavior and RNNs for modeling history dependence (non-equilibrium). This section also describes the training dataset generation, the incorporation of thermodynamic constraints, and the architectural design of the ML models. In Section 4, we evaluate the proposed model through a series of benchmark cases involving bulk and isochoric deformations, varying loading conditions, different stretch levels, thermodynamic adherence, and noise sensitivity. Finally, in Section 5, we draw conclusions.

## 2 Generalized internal state variable-based visco-hyperelastic constitutive framework

### 2.1 Basic Kinematics

A soft material transitions from its initial undeformed configuration $\Omega_0$ at time $t = 0$, to a deformed configuration $\Omega$ at time $t$. The macroscopic motion over time maps the reference position vector **X** to a new spatial position vector **x** through a vector function of $\mathbf{x} = \chi(\mathbf{X}, t) \in \Omega$. This is described by the deformation gradient tensor as $d\mathbf{x} = \mathbf{F}(\mathbf{X}, t) d\mathbf{X}$, ($\det \mathbf{F}(\mathbf{X}, t) = J(\mathbf{X}, t)$), where $J$ quantifies the volume change. In the Lagrangian framework, the deformation is characterized by using the symmetric right Cauchy-Green deformation tensor $\mathbf{C} = \mathbf{F}(\mathbf{X}, t)^T \mathbf{F}(\mathbf{X}, t)$. To address volumetric and isochoric deformations in soft materials, **F** can be multiplicatively split into a dilational component $J^{1/3}\mathbf{I}$ and a distortional component $\overline{\mathbf{F}}$. From this decomposition, the modified right Cauchy-Green tensor ($\overline{\mathbf{C}}$) yields $\mathbf{C} = (J^{2/3}\mathbf{I})\overline{\mathbf{C}}$, in which $\det \overline{\mathbf{C}} = (\det \overline{\mathbf{F}})^2 = 1$.

To allow for a more systematic treatment of the material's response, the Helmholtz free energy function $\Psi$ (defined per unit reference volume) can be decoupled into separate volumetric and isochoric/distortional contributions. In the generalized internal state variable-based visco-hyperelastic constitutive framework, the isochoric component of the Helmholtz free energy is further split into a part that describes the thermodynamic equilibrium state as time $t \to \infty$, and a set of configurational free energies $\Upsilon_\alpha, \alpha = 1, \ldots, m$, responsible for the thermodynamic non-equilibrium state of the material. The scalar-valued functions $\Upsilon_\alpha$ are characterized by a set of internal variables $\Gamma_\alpha$. Overall, the total free energy is given as

$$\Psi(\mathbf{C}, \Gamma_1, \ldots, \Gamma_m) = \Psi_{\text{VOL}}^\infty(J) + \Psi_{\text{ISO}}^\infty(\overline{\mathbf{C}}) + \sum_{\alpha=1}^{m} \Upsilon_\alpha(\overline{\mathbf{C}}, \Gamma_\alpha) \tag{1}$$

where $\psi_{vol}^\infty$ and $\psi_{iso}^\infty$ are the equilibrium volumetric and isochoric free energy components. A key principle of thermodynamics, Clausius-Duhem inequality represents a fundamental expression of the second law and can be used to compute internal dissipation $D_{\text{int}}$ during deformation. For isothermal processes, when considering the second Piola-Kirchhoff stress tensor **S**, the internal dissipation $D_{\text{int}}$ can be as $D_{\text{int}} = \mathbf{S} : \dot{\mathbf{C}}/2 - \dot{\Psi} \geq 0 \ \forall t \in [0, T]$ where $\dot{\Psi}$ is $d\Psi/dt$. **S** is thermodynamically conjugate to $\dot{\mathbf{C}}$, the rate of change of the right Cauchy-Green deformation tensor through the Coleman-Noll procedure, which imposes constraints on the constitutive equations to reach $\mathbf{S} = 2 \partial\Psi/\partial\mathbf{C}$ for compressible hyperelasticity. Using **Eq. 1** and applying the chain rule,

$$\dot{\Psi} = \frac{\partial \Psi_{\text{VOL}}^{\infty}(J)}{\partial J}\dot{J} + \left[\left(\frac{\partial \Psi_{\text{ISO}}^{\infty}(\overline{\mathbf{C}})}{\partial \overline{\mathbf{C}}} + \sum_{\alpha=1}^{m}\frac{\partial \Upsilon_{\alpha}(\overline{\mathbf{C}},\boldsymbol{\Gamma}_{\alpha})}{\partial \overline{\mathbf{C}}}\right):\dot{\overline{\mathbf{C}}} + \sum_{\alpha=1}^{m}\frac{\partial \Upsilon_{\alpha}(\overline{\mathbf{C}},\boldsymbol{\Gamma}_{\alpha})}{\partial \boldsymbol{\Gamma}_{\alpha}}:\dot{\boldsymbol{\Gamma}}_{\alpha}\right] \tag{2}$$

By taking the derivative of $\dot{\overline{\mathbf{C}}}$ with respect to the symmetric tensor $\mathbf{C}$, $\dot{\overline{\mathbf{C}}} = \partial \overline{\mathbf{C}}/\partial \mathbf{C}:\dot{\mathbf{C}}$, along with the first derivative of $J$, $\dot{J} = \partial J/\partial \mathbf{C}:\dot{\mathbf{C}} = J\mathbf{C}^{-1}:\dot{\mathbf{C}}/2$ the internal dissipation yields:

$$D_{\text{int}} = \left[\mathbf{S} - J\frac{\partial \Psi_{\text{VOL}}^{\infty}(J)}{\partial J}\mathbf{C}^{-1} - 2\left(\frac{\partial \Psi_{\text{ISO}}^{\infty}(\overline{\mathbf{C}})}{\partial \mathbf{C}} + \sum_{\alpha=1}^{m}\frac{\partial \Upsilon_{\alpha}(\overline{\mathbf{C}},\boldsymbol{\Gamma}_{\alpha})}{\partial \mathbf{C}}\right)\right]:\frac{1}{2}\dot{\mathbf{C}} - \sum_{\alpha=1}^{m}\frac{\partial \Upsilon_{\alpha}(\overline{\mathbf{C}},\boldsymbol{\Gamma}_{\alpha})}{\partial \boldsymbol{\Gamma}_{\alpha}}:\dot{\boldsymbol{\Gamma}}_{\alpha} \geq 0 \tag{3}$$

Regarding $\mathbf{S} = 2\nabla_{\mathbf{C}}\Psi$ and Coleman-Noll procedure, the total stress can be decoupled into two equilibrium components, volumetric ($\mathbf{S}_{\text{VOL}}^{\infty}$) and isochoric hyperelastic ($\mathbf{S}_{\text{ISO}}^{\infty}$), and a non-equilibrium component, $\sum_{\alpha=1}^{m}\mathbf{Q}_{\alpha}$, such that

$$\mathbf{S} = \mathbf{S}_{\text{ISO}}^{\infty} + \mathbf{S}_{\text{VOL}}^{\infty} + \sum_{\alpha=1}^{m}\mathbf{Q}_{\alpha} = J\frac{\partial \Psi_{\text{VOL}}^{\infty}(J)}{\partial J}\mathbf{C}^{-1} + 2\frac{\partial \Psi_{\text{ISO}}^{\infty}(\overline{\mathbf{C}})}{\partial \mathbf{C}} + \sum_{\alpha=1}^{m}2\frac{\partial \Upsilon_{\alpha}(\overline{\mathbf{C}},\boldsymbol{\Gamma}_{\alpha})}{\partial \mathbf{C}} \tag{4}$$

The stress components $\mathbf{S}_{\text{VOL}}^{\infty}$ and $\mathbf{S}_{\text{ISO}}^{\infty}$ are derived from equilibrium thermodynamic potentials and are fully recoverable, meaning they do not result in entropy production or dissipation during deformation. The only source of dissipation arises from the non-equilibrium configurational stresses $\sum_{\alpha=1}^{m}\mathbf{Q}_{\alpha}$, which are associated with time-dependent internal variables representing the material's viscous behavior. Thus, a non-negative inequality as well as a constitutive framework may be obtained as

$$\begin{aligned}
D_{\text{int}} &= -\sum_{\alpha=1}^{m}\frac{\partial \Upsilon_{\alpha}(\overline{\mathbf{C}},\boldsymbol{\Gamma}_{\alpha})}{\partial \boldsymbol{\Gamma}_{\alpha}}:\dot{\boldsymbol{\Gamma}}_{\alpha} \geq 0 \\
\mathbf{S}_{\text{VOL}}^{\infty} &= J\frac{\partial \Psi_{\text{VOL}}^{\infty}(J)}{\partial J}\mathbf{C}^{-1}, \\
\mathbf{S}_{\text{ISO}}^{\infty} &= 2\frac{\partial \Psi_{\text{ISO}}^{\infty}(\overline{\mathbf{C}})}{\partial \mathbf{C}}, \\
\mathbf{Q}_{\alpha} &= 2\frac{\partial \Upsilon_{\alpha}(\overline{\mathbf{C}},\boldsymbol{\Gamma}_{\alpha})}{\partial \mathbf{C}}, \\
\mathbf{S}_{\text{ISO}} &= \mathbf{S}_{\text{ISO}}^{\infty} + \sum_{\alpha=1}^{m}\mathbf{Q}_{\alpha},
\end{aligned} \tag{5}$$

Note, $\sum_{\alpha=1}^{m} \mathbf{Q}_\alpha$ represents the irreversible kinetic relations in which a series of tensor variables are associated with the non-equilibrium Maxwell branches. Modeling this nonequilibrium response of viscoelastic materials requires two key components: a rheological model and an evolution equation. The rheological model, such as the generalized Maxwell model (**Figure 1**), captures both instantaneous and time-dependent responses, while the evolution equation describes the time-dependent behavior of internal variables under loading.

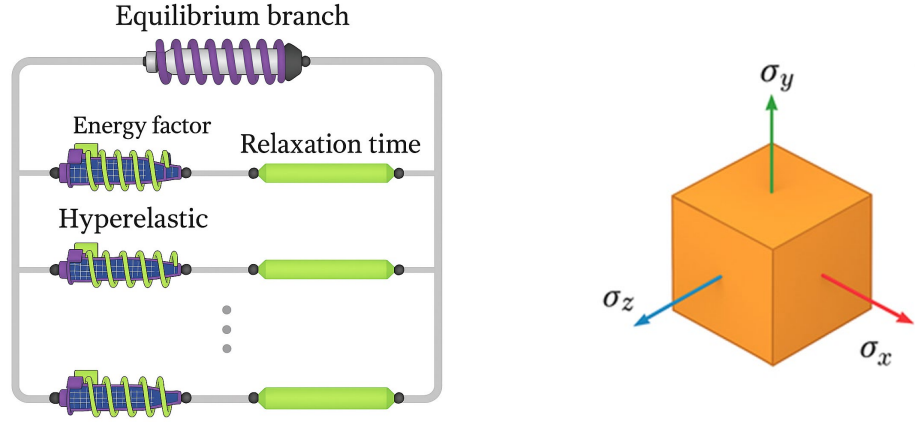

a) Nonlinear Viscous Parts in the Generalized Maxwell Model,   b) Multiaxial Loading on Material Point

Figure 1: Modeling framework combining viscoelastic behavior and multiaxial stress conditions.

### 2.1.1 Equilibrium:

According to **Eq.5$_b$** and **5$_c$**, and assuming isochoric components of the second Piola–Kirchhoff stress in the equilibrium branch has continuous derivatives for the determinant of deformation gradient J as well as principal invariants $I_1$, and $I_2$, the total equilibrium stress is given as

$$\mathbf{S}^\infty = \frac{\partial \Psi^\infty_{\text{VOL}}(J)}{\partial J}\frac{\partial J}{\partial \mathbf{C}} + 2\sum_{k=1}^{2}\frac{\partial \Psi^\infty_{\text{ISO}}(\overline{\mathbf{C}})}{\partial \overline{I}_k}\frac{\partial \overline{I}_k}{\partial \mathbf{C}} \tag{6}$$

Considering the second equilibrium stress component, we know $\dfrac{\partial \overline{\mathbf{C}}}{\partial \mathbf{C}} = J^{-2/3}\mathbb{P}^T$ in which $\mathbb{P}^T$ is the projection tensor ($\mathbb{I} - \dfrac{1}{3}\mathbf{C}\otimes\mathbf{C}^{-1}$), so that $\mathbb{P}:\mathbf{Z} = \text{Dev}(\mathbf{Z})$ ([17]). Therefore, we have:

$$\left(\frac{\partial \overline{\mathbf{C}}}{\partial \mathbf{C}}\right)^T : \frac{\partial \overline{I}_k}{\partial \overline{\mathbf{C}}} = J^{-2/3}\text{Dev}\left(\frac{\partial \overline{I}_k}{\partial \overline{\mathbf{C}}}\right) \tag{7}$$

Considering **Eq.7** and $\frac{\partial J}{\partial \mathbf{C}} = \frac{1}{2}J\mathbf{C}^{-1}$, as well as the derivatives of the first and second invariants, $\frac{\partial I_1}{\partial \mathbf{C}} = \mathbf{I}$ and $\frac{\partial I_2}{\partial \mathbf{C}} = I_1\mathbf{I} - \mathbf{C}$, we can rewrite **Eq. 6** as:

$$\mathbf{S}^{\infty} = J\frac{\partial \Psi_{VOL}^{\infty}(J)}{\partial J}\mathbf{C}^{-1} + 2J^{-2/3}\left[\frac{\partial \Psi_{ISO}^{\infty}(\overline{\mathbf{C}})}{\partial \overline{I}_1}\text{Dev}\left(\frac{\partial \overline{I}_1}{\partial \overline{\mathbf{C}}}\right) + \frac{\partial \Psi_{ISO}^{\infty}(\overline{\mathbf{C}})}{\partial \overline{I}_2}\text{Dev}\left(\frac{\partial \overline{I}_2}{\partial \overline{\mathbf{C}}}\right)\right] \quad (8)$$

*2.1.2 Evolution of the non-equilibrium stress components:*

The evolution equation mathematically links stress, strain rate, and deformation history while adhering to the entropy inequality. Although the Boltzmann superposition principle is often used, it may not accurately capture the behavior of nonlinear materials under high stresses or large deformations. As an alternative, various differential equation-based frameworks have been developed to handle nonlinear responses and energy dissipation mechanisms ([17], [120]). The associated dissipative evolution equations follow a well-established classical generalized form:

$$\dot{\mathbf{Q}}_\alpha = \hat{f}(\overline{\mathbf{C}}, \boldsymbol{\Gamma}_1, ..., \boldsymbol{\Gamma}_\alpha) \quad \alpha = 1,...,m \quad (9)$$

where $\hat{f}$ is a constitutive function that governs the evolution of internal variables based on the deformation history and internal state.

## 2.2 An alternative form of the generalized visco-hyperelastic constitutive framework

Recently, Upadhyay et al. demonstrated [106] that the individual stress components in the generalized external state variable-based visco-hyperelastic constitutive framework can be written as a weighted linear combination of the components of certain irreducible integrity bases. It was demonstrated that ML mappings between strain/strain rate invariants and the coefficients of the integrity basis components (referred to as response functions) enforce specific physics-based constraints, including the principles of local action and determinism, balance of angular momentum, objectivity, and an assumed isotropic material symmetry. Such an ML mapping, based on invariants and response functions instead of raw stress-strain pairs, also promotes data efficiency, model generalizability to unseen deformation states (those considered during model training), and improved model validation as the learned response functions can be conveniently compared with known theoretical models. Following the above approach, we expand the generalized stress components in **Eq.8** via the chain rule, leading to the following alternative form of the generalized internal state variable-based visco-hyperelastic constitutive model:

$$\mathbf{S}_{\text{VOL}}^{\infty} = \delta(J)\mathbf{C}^{-1} \tag{10}$$

$$\mathbf{S}_{\text{ISO}}^{\infty} = J^{-2/3}\left[\chi_1(\bar{I}_1,\bar{I}_2)\text{Dev}(\mathbf{I}) + \chi_2(\bar{I}_1,\bar{I}_2)\text{Dev}(\bar{\mathbf{C}})\right]$$

Inspired by the integrity basis representation in the equilibrium stress **Eq. 10b** and the representation of each non-equilibrium branch evolution through **Eq. 9**, the sum of all non-equilibrium branches can also be expressed using the same integrity basis:

$$\sum_{\alpha=1}^{m}\mathbf{Q}_\alpha = J^{-2/3}\left[\sum_{\alpha=1}^{m}\xi_{1,\alpha}(t,\bar{I}_1,\bar{I}_2)\text{Dev}(\mathbf{I}) + \sum_{\alpha=1}^{m}\xi_{2,\alpha}(t,\bar{I}_1,\bar{I}_2)\text{Dev}(\bar{\mathbf{C}})\right] \tag{11}$$

Note that the integrity basis components of the total stress S are

$$\mathbb{G}_1 = \mathbf{C}^{-1}, \quad \mathbb{G}_2 = \text{Dev}(\mathbf{I}), \quad \mathbb{G}_3 = \text{Dev}(\bar{\mathbf{C}}) \tag{12}$$

and the response functions are given by

$$\delta(J) = J\frac{\partial \Psi_{\text{VOL}}^{\infty}(J)}{\partial J} \quad (a) \tag{13}$$

$$\chi_1(\bar{I}_1,\bar{I}_2) = 2\left(\frac{\partial \Psi_{\text{ISO}}^{\infty}(\bar{I}_1,\bar{I}_2)}{\partial \bar{I}_1} + \bar{I}_1\frac{\partial \Psi_{\text{ISO}}^{\infty}(\bar{I}_1,\bar{I}_2)}{\partial \bar{I}_2}\right), \quad (b)$$

$$\chi_2(\bar{I}_1,\bar{I}_2) = -2\frac{\partial \Psi_{\text{ISO}}^{\infty}(\bar{I}_1,\bar{I}_2)}{\partial \bar{I}_2} \quad (c)$$

$$\xi_{1,\alpha}(t,\bar{I}_1,\bar{I}_2) = 2\left(\frac{\partial \Upsilon_\alpha(t,\bar{I}_1,\bar{I}_2)}{\partial \bar{I}_1} + \bar{I}_1\frac{\partial \Upsilon_\alpha(t,\bar{I}_1,\bar{I}_2)}{\partial \bar{I}_2}\right), \quad (d)$$

$$\xi_{2,\alpha}(t,\bar{I}_1,\bar{I}_2) = -2\frac{\partial \Upsilon_\alpha(t,\bar{I}_1,\bar{I}_2)}{\partial \bar{I}_2} \quad (e)$$

Notice that the total stress in this alternative form of the generalized visco-hyperelastic constitutive framework is a linear combination of integrity basis components multiplied by their coefficient response functions, and so the individual stress components can be represented as the following system of equations:

$$\left[ \text{vec}\left(\mathbf{S}_{\text{VOL}}^{\infty}\right) \right] = \left[ \text{vec}(\mathbb{G}_1) \right]\left[\delta(J)\right] \tag{14}$$

$$\left[ \text{vec}\left(\frac{\mathbf{S}_{\text{ISO}}^{\infty}}{J^{-2/3}}\right) \right] = \left[ \text{vec}(\mathbb{G}_2) \quad \text{vec}(\mathbb{G}_3) \right] \begin{bmatrix} \chi_1(\overline{I}_1, \overline{I}_2) \\ \chi_2(\overline{I}_1, \overline{I}_2) \end{bmatrix}$$

$$\left[ \text{vec}\left(\frac{\sum_{\alpha=1}^{m} \mathbf{Q}_\alpha}{J^{-2/3}}\right) \right] = \left[ \text{vec}(\mathbb{G}_2) \quad \text{vec}(\mathbb{G}_3) \right] \begin{bmatrix} \sum_{\alpha=1}^{m} \xi_{1,\alpha}(t, \overline{I}_1, \overline{I}_2) \\ \sum_{\alpha=1}^{m} \xi_{2,\alpha}(t_1, \overline{I}, \overline{I}_2) \end{bmatrix}$$

where vec(.) is the matrix Voigt form. The above equations are equivalent to **Eq. 5** and are expressed in the form [b]=[A][x]. The vector [x], which contains the functional coefficients (i.e., response functions), enables the extraction of these coefficients through a least-squares minimization approach ( $min \| [A][x] - [b] \|_2^2$ ).

The above system of equations describes different contributions to stress responses in an isotropic, history-dependent visco-hyperelastic material. The first equation, with the $\delta(J)$ response function, represents the volumetric stress contribution, ensuring proper compressibility behavior. The next equation, containing response functions $\chi_1(\overline{I}_1, \overline{I}_2)$ and $\chi_2(\overline{I}_1, \overline{I}_2)$, defines the time-independent equilibrium isochoric stress component. The remaining equation, containing response functions $\xi_{1,\alpha}(t, \overline{I}_1, \overline{I}_2)$, $\xi_{2,\alpha}(t, \overline{I}_1, \overline{I}_2)$ incorporates isochoric, non-equilibrium response, which depends on the evolution of internal variables. These terms modify the stress response by incorporating internal state variables, allowing for the modeling of time-dependent or path-dependent deformations. Being a function of time ensures that the material's isochoric response dynamically adapts to deformation history, reflecting relaxation and creep. Note that the integrity basis of tensors is beneficial in ML-based material modeling because it provides a minimal and complete set of scalar invariants to describe the dependence of a function on one or more tensors. From **Eq. 13**, the response functions are functions of the strain invariants. In available internal state variable-based visco-hyperelastic constitutive models, these functions as well as the evolution equation of the non-equilibrium isochoric stress component take explicit mathematical forms. Unlike these rigid models, the physics-informed data-driven constitutive model proposed in this study will discover the mapping between invariants and response functions and the evolution of the time-dependent response functions directly from the experimental data.

# 3 Proposed physics-informed data-driven constitutive model

## 3.1 Data-Driven Mapping

Based on **Eq. 14**, we define three machine learning mappings, one for each of the volumetric, isochoric hyperelastic, and isochoric viscous stress components, which comprise our physics-informed data-driven constitutive model:

1. The volumetric stress component surrogate model $\tilde{M}_{vol}$ (**Eq. 15a**) learns the response function $\delta(J)$ and captures how compressibility affects volumetric stress using a bulk energy function.

2. The isochoric hyperelastic stress surrogate model $\tilde{M}_{iso}$ (**Eq. 15b**) takes the principal invariants as inputs and learns the response functions $\chi_1(\bar{I}_1, \bar{I}_2)$ and $\chi_2(\bar{I}_1, \bar{I}_2)$, which describes the material's deviatoric stress response.

3. The isochoric viscoelastic stress surrogate model $\tilde{M}_{visco}$ (**Eq. 15c**) incorporates time-dependent effects by learning $\sum_{\alpha=1}^{m} \xi_{1,\alpha}(t, \bar{I}_1, \bar{I}_2)$ and $\sum_{\alpha=1}^{m} \xi_{2,\alpha}(t, \bar{I}_1, \bar{I}_2)$. These response functions track the evolution of stress over time, capturing stress components.

$$\tilde{M}_{vol} : J \in \mathbb{R}^1 \to \delta \in \mathbb{R}^1 \tag{15}$$

$$\tilde{M}_{iso} : (I_1, I_2) \in \mathbb{R}^2 \to \chi \in \mathbb{R}^2,$$

$$\tilde{M}_{visco} : (t, I_1, I_2) \in \mathbb{R}^3 \to \sum \xi_{,\alpha} \in \mathbb{R}^2$$

Each surrogate model can be trained on a unique dataset (the next subsection describes training dataset generation), structured to capture the relevant dependencies and ensure accurate predictions. For the volumetric stress surrogate model $\tilde{M}_{vol}$, the dataset is based on the volume ratio J, with the corresponding stress response function $\delta(J)$ as the target variable. This function governs the volumetric response and can be used to calculate the volumetric stress component via **Eq. 10a**. Similarly, for the isochoric hyperelastic stress surrogate model $\tilde{M}_{iso}$, the dataset is based on the input variables $\bar{I}_1$ and $\bar{I}_2$ with the corresponding response functions $\chi_1$ and $\chi_2$ as target variables, which, in turn, can be used to compute the isochoric hyperelastic stress component using **Eq. 10b**. Lastly, for the isochoric viscoelastic stress surrogate model $\tilde{M}_{visco}$ accounts for the history-dependent nature of stress, where relaxation terms are learned as a function of time $t$, stretch invariants $\bar{I}_1$, and $\bar{I}_2$. The time evolution of the target variables $\sum_{\alpha=1}^{m} \xi_{1,\alpha}$, and $\sum_{\alpha=1}^{m} \xi_{2,\alpha}$ which collectively represent the contribution of all Maxwell viscoelastic branches, effectively captures the material's transient mechanical response. Using these target variables, the viscoelastic stress component can be calculated using **Eq. 11**. Overall, with the trained set of surrogate models, the three stress components can be calculated as:

$$\tilde{\mathbf{S}}_{\text{VOL}}^{\infty,i} = \tilde{\delta}^i(J)\mathbb{G}_1^i \tag{16}$$

$$\tilde{\mathbf{S}}_{\text{ISO}}^{\infty,j} = J^{-2/3}\left[\tilde{\chi}_1^j(\overline{I}_1,\overline{I}_2)\mathbb{G}_2^j + \tilde{\chi}_2^j(\overline{I}_1,\overline{I}_2)\mathbb{G}_3^j\right]$$

$$\sum_{\alpha=1}^{m}\tilde{\mathbf{Q}}_\alpha^k = J^{-2/3}\left[\sum_{\alpha=1}^{m}\tilde{\xi}_{1,\alpha}^k(t,\overline{I}_1,\overline{I}_2)\mathbb{G}_2^k + \sum_{\alpha=1}^{m}\tilde{\xi}_{2,\alpha}^k(t,\overline{I}_1,\overline{I}_2)\mathbb{G}_3^k\right]$$

where $\tilde{\bullet}$ denotes the approximated integrity basis values of the stresses and coefficients predicted through the surrogate models.

## 3.2 Training Dataset Generation

In general, deformation measurements and the corresponding total stress response of the material can be obtained through experiments and/or simulations. These measurements provide a direct representation of the material's macroscopic behavior but do not inherently separate the contributions from volumetric, isochoric, and viscoelastic effects. Thus, defining datasets that clearly distinguish between the raw mechanical response dataset $\mathcal{D}$ (comprising stresses, strains, and strain rates) and the processed or alternative dataset $\mathcal{D}^*$ (comprising time, invariants, and the response functions) to train the ML models is essential. To make the learning process both efficient and consistent with physical laws, the data must be reformulated in terms of scalar invariants and basis coefficients; this is what $\mathcal{D}^*$ can provide for the ML modeling. Computed invariants serve as the primary features for training data-driven models for stress prediction; thus, we define an $\mathcal{D}^*$ as

$$\mathcal{D}^* \coloneqq \left\{J^i, \delta(J^i)\right\}_{i=1}^{N_{vol}} \cup \tag{17}$$

$$\left\{\left[\overline{I}_1^j,\overline{I}_2^j\right],[\chi_1(\overline{I}_1^j,\overline{I}_2^j),\chi_2(\overline{I}_2^j)]\right\}_{j=1}^{N_{iso}} \cup$$

$$\left\{\left[t^k,\overline{I}_1^k,\overline{I}_2^k\right],[\xi_{1,\alpha}(t^k,\overline{I}_1^k,\overline{I}_2^k),\xi_{2,\alpha}(t^k,\overline{I}_2^k)]\right\}_{k=1}^{N_{visco}}$$

Since the ML model is built on stress decomposition for training and prediction, it is essential to develop methodologies that can separate these stress components, ensuring consistency between theoretical datasets and real-world measurements, particularly in applications where time-dependent effects influence long-term performance. The following subsection discusses a robust approach to achieving this goal.

### 3.2.1 Methodology for Decomposing Total Stress

Generally, experiments are conducted at different strain rates to extract the mechanical behavior in viscoelastic materials (**Figure 2**). To allow for analysis of the relaxation behavior and identification of the long-term stress equilibrium, the total stress can be extracted at a specific constant strain (stretch) level for each test and then plotted as a function of time. By fitting a suitable function to the stress relaxation curve, we determine its asymptotic stress level, which corresponds to the total stress in the quasi-static/equilibrium regime ($\mathbf{S}^\infty = \mathbf{S}_{vol}^\infty + \mathbf{S}_{iso}^\infty$). This asymptotic analysis is critical in confirming that the viscoelastic contribution has fully relaxed, ensuring that only the intrinsic volumetric and isochoric hyperelastic stress components remain, and the non-equilibrium viscoelastic stress components become insignificant. Then, the non-equilibrium stress component can be obtained by subtracting the resolved quasi-static stress from the total stress at a given strain rate (i.e., $\sum_{\alpha=1}^{m} \mathbf{Q}_\alpha = \mathbf{S}_{total} - \mathbf{S}^\infty$).

Moreover, to obtain a volumetric stress component, we can also consider a hydrostatic test, where the material is subjected to uniform compression in all directions. Since no shear deformation occurs in this test, the measured stress corresponds exclusively to the bulk response, isolating $\mathbf{S}_{vol}^\infty$.

Once the volumetric stress component is determined, the equilibrium isochoric stress is obtained by subtracting from the total equilibrium stress measured under general loading conditions from the asymptotic analysis ($\mathbf{S}_{iso}^\infty = \mathbf{S}^\infty - \mathbf{S}_{vol}^\infty$). Since viscoelastic effects have already dissipated in the quasi-static regime, the extracted $\mathbf{S}_{iso}^\infty$ represents the proper equilibrium isochoric stress component.

The above procedure can be repeated at every stretch value to resolve the three stress components ($\mathbf{S}_{vol}^\infty, \mathbf{S}_{iso}^\infty$, and $\sum_{\alpha=1}^{m} \mathbf{Q}_\alpha$) as a function of time and $\mathbf{C}$ as well as its invariants and the integrity basis components (**Eq. 12**). Then, using the system of linear equations in **Eq. 14**, the corresponding response functions of the three stress components can be computed, thus finalizing the dataset $\mathcal{D}^*$ that can be directly utilized for training the data-driven surrogate models (**Eq. 15**).

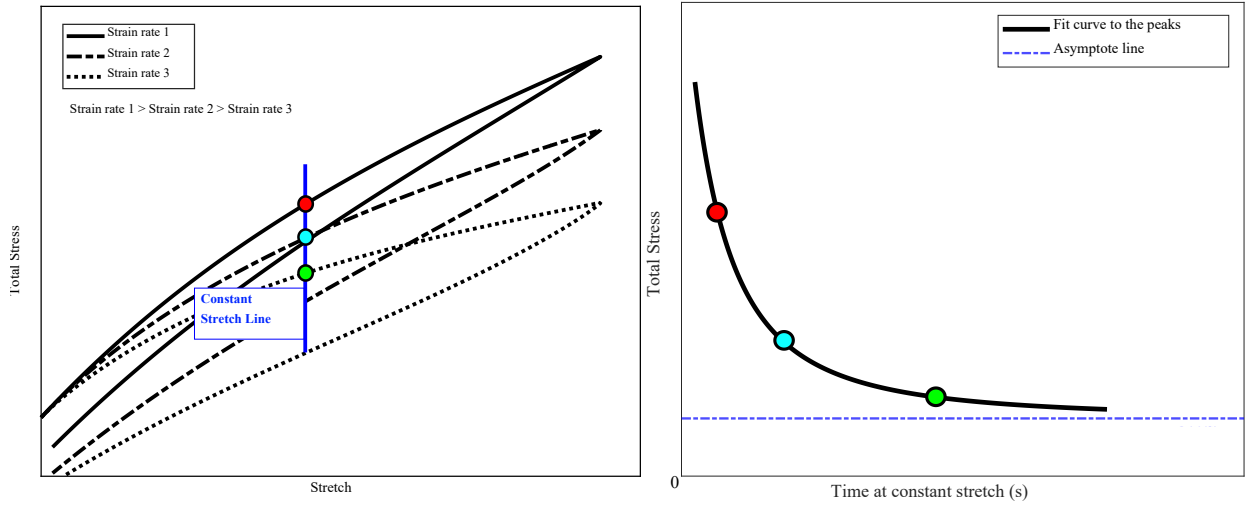

a) Stress-Stretch curves at different strain rates     b) Strain rate curve to find asymptote line

Figure 2: Schematic of different cyclic experimental tests for viscoelastic materials

## 3.3 Physics-Based Constraints on the Data-Driven Constitutive Model

Owing to the particular mappings of the proposed data-driven constitutive model (**Eq. 15**) and the associated stress-prediction equations (**Eq. 16**), a number of physics-based restrictions are intrinsically imposed:

- *Objectivity (Frame-Indifference)*

    The strain energy function must be independent of rigid body motions: $\Psi(\mathbf{C}) = \Psi(\mathbf{Q}\mathbf{C}\mathbf{Q}^T), \forall \mathbf{Q} \in SO(3)$. The proposed model fulfills this constraint as **C**, and thus, the integrity basis components in **Eq. 16** are objective tensors. The material response does not change under rotations, maintaining consistency in different reference frames.

- *Material Symmetry*

    For isotropic materials, $\Psi(\mathbf{C})$ should be invariant under the symmetry group: $\Psi(\mathbf{C}) = \Psi(\mathbf{Q}^T\mathbf{C}\mathbf{Q}), \forall \mathbf{Q} \in SO(3)$. The proposed model imposes isotropic symmetry owing to the incorporation of isotropic principal invariants $I_1$, $I_2$, and $J$ (**Eq.15**).

- *Balance of Angular Momentum*

    **S** must be symmetric $\mathbf{S}=\mathbf{S}^T$. The proposed model imposes the balance of angular momentum constraint as the integrity basis components in **Eq. 16** are necessarily

symmetric. Thus, the second Piola-Kirchhoff stress tensor (and thus, the Cauchy stress tensor as well) remains symmetric, avoiding unphysical internal torques.

However, we need to incorporate two additional constraints in our ML surrogate models:

- Stress-free reference state of equilibrium stress: This constraint requires that at **C=I**, $\mathbf{S}_{vol}^{\infty} = \mathbf{S}_{iso}^{\infty} = \mathbf{0}$. This condition will be imposed by using GPR for the $\tilde{M}_{vol}$ and $\tilde{M}_{iso}$ surrogate models as described later.
- The second law of thermodynamics: This constraint requires that the dissipation by non-equilibrium stress components is non-negative (**Eq. 5a**). This constraint will be imposed in the memory-dependent RNN, which is used in this work for the $\tilde{M}_{visco}$ surrogate model.

To summarize, we employ a hybrid machine learning approach to capture both hyperelastic and viscoelastic material behaviors: a GPR algorithm to learn the equilibrium volumetric and isochoric strain energy functions, and an RNN model to learn the time-dependent viscoelastic response.

### 3.4 Gaussian Process Regression for Hyperelasticity

To accurately represent hyperelastic response which can be decomposed into equilibrium volumetric and isochoric components, we employ GPR with a stress-free reference statement embedded into the kernel structure. The GPR-supervised learning method is used for regression tasks. In this study, the Matérn 3/2 kernel (Matérn, 1960 [121]) is used to define the covariance structure, and the optimal parameters are determined using the maximum log-likelihood approach. For a detailed discussion on GPR, please refer to [106],[122].

Once trained, GPR predicts outputs for new inputs using the learned kernel function. In this study, GPR is applied to develop surrogate models for the volumetric ($\tilde{M}_{vol}$) and isochoric ($\tilde{M}_{iso}$) stress components. To ensure that the model satisfies the stress-free reference configuration, a normalization condition is enforced, guaranteeing that the equilibrium volumetric and isochoric stress components vanish in the undeformed configuration. This is achieved by incorporating stress-free reference configuration into the training dataset:

$$\mathbf{S}_{\text{VOL/ISO}}^{\infty}(\mathbf{C=I}) = \mathbf{0} \tag{18}$$

By appropriately tuning the noise parameter (which controls the strength of the Tikhonov regularization), the model can satisfy the physical requirement of zero equilibrium stress in the undeformed state. The GPR models separately learn the volumetric response function $\delta(J)$ as a function of $J$ and the isochoric response function coefficients $\chi_1(\bar{I}_1, \bar{I}_2)$ and $\chi_2(\bar{I}_1, \bar{I}_2)$ as a function of invariants $\bar{I}_1$ and $\bar{I}_2$. As mentioned before, the reference state of $J = 1$ and $I_1 = I_2 = 3$, along with $\delta = 0$ and $\chi_1 = \chi_2 = 0$, must be considered in the training dataset to enforce a stress-free reference state.

## 3.5 RNN for history-dependent response

3.5.1 Vanilla and LSTM networks

RNNs are designed to store past information using hidden state vectors, making them ideal for processing sequential data. A basic RNN model, commonly referred to as a vanilla RNN, is illustrated in **Figure 3a**. Here, the hidden state $h^{(t-1)}$ at the previous time step $t_{n-1}$ stores historical data, while $x^{(t)}$ and $y^{(t)}$ represent the input and output at the current time step $t_n$, respectively. The model updates its hidden state and outputs as follows:

$$h^{(t)} = A\left(\Omega_{hh} h^{(t-1)} + \Omega_{hx} x^{(t)} + B_h\right) \tag{19}$$

$$y^{(t)} = A\left(\Omega_{yh} h^{(t)} + B_y\right) \tag{20}$$

where $\Omega_{hh}$ and $\Omega_{hx}$ are weight matrices that control how past hidden states and the new input influence the current hidden state. $\Omega_{yh}$ is a weight matrix to control how the hidden state influences the output. $B_h$ and $B_y$ are biased terms. Finally, $A(\cdot)$ is an activation function (commonly tanh or sigmoid) that introduces non-linearity to the mapping and ensures stability in learning.

At each time step t, the RNN model takes an input $x(t)$ and combines it with the previous hidden state $h^{(t-1)}$ using **Eq. 19**. After updating the hidden state $h^{(t)}$, the network computes the output at time t using **Eq. 20**. For the RNN to learn from data, it must optimize its parameters. This may be done by minimizing its loss function $L$. The training process involves computing the gradients of the loss function to the parameters and updating them using gradient descent. Since RNNs operate over sequences, gradients must be computed over multiple time steps, a process known as Back Propagation Through Time (BPTT) [111]. The weight update equation for $\Omega_{yh}$, $\Omega_{hh}$, and $\Omega_{hx}$ can be expressed as

$$\Omega_{\kappa} = \Omega_{\kappa} - \eta \frac{\partial L}{\partial \Omega_{\kappa}}, \quad \kappa = yh, hh, hx \tag{21}$$

where $\eta$ denotes the learning rate that controls the step size of the update. This process is repeated for many iterations until the model learns the best values for $\Omega_{yh}$, $\Omega_{hh}$, and $\Omega_{hx}$. The gradient $\partial L / \partial \Omega_{\kappa}$ tells how much to change the weight to minimize the loss function, and it accumulates over all time steps. However, when sequences are long, gradients tend to shrink exponentially due to repeated multiplication, leading to the vanishing gradient problem. More advanced architectures, such as LSTM networks [112] has been developed to overcome these issues. LSTMs solve the vanishing gradient problem by introducing a memory cell $c^{(t)}$ that stores information across long sequences (**Figure 3b**). Three key gates control this memory:

the forget gate, the input gate, and the output gate. Using **Eq. 22**, the Forget gate $f^{(t)}$, $\kappa = f^{(t)}$, decides whether to retain or discard past information. If $f^{(t)} = 1$, memory is preserved; if the Output gate $\kappa = o^{(t)} = 0$, it is erased. Next, the Input gate $i^{(t)}$, ($\kappa = i^{(t)}$ in **Eq. 22**) determines what new information should be added to the cell state. The candidate memory $\tilde{c}^{(t)}$ is computed using **Eq. 23** and an activation function A(.), which outputs values between -1 and 1, ensuring numerical stability. To update the cell state $c^{(t)}$, the forget gate first scales the previous memory $c^{(t-1)}$, and then the input gate determines how much of $\tilde{c}^{(t)}$ should be incorporated ($c^{(t)} = f^{(t)} c^{(t-1)} + i^{(t)} \tilde{c}^{(t)}$). Finally, the Output gate $o^{(t)}$, ($\kappa = o^{(t)}$ in **Eq. 22**) controls how much of the updated cell state should influence the hidden state [111]. The hidden state is then computed as: $h^{(t-1)} = o^{(t)} A(c^{(t)})$.

$$\kappa^{(t)} = \sigma\left(\Omega_{\kappa a} a^{(t-1)} + \Omega_{\kappa x} x^{(t)} + b_\kappa\right) \quad (22)$$

$$\tilde{c}^{(t)} = A\left(\Omega_{ca} h^{(t-1)} + \Omega_{cx} x^{(t)} + B_c\right) \quad (23)$$

If $o^{(t)} = 1$, the hidden state receives strong influence from $c^{(t)}$, and if $o^{(t)} = 0$, minimal information is passed forward. This gating mechanism ensures that only relevant information is passed to the next step, allowing LSTMs to effectively capture long-term dependencies without suffering from vanishing gradients.

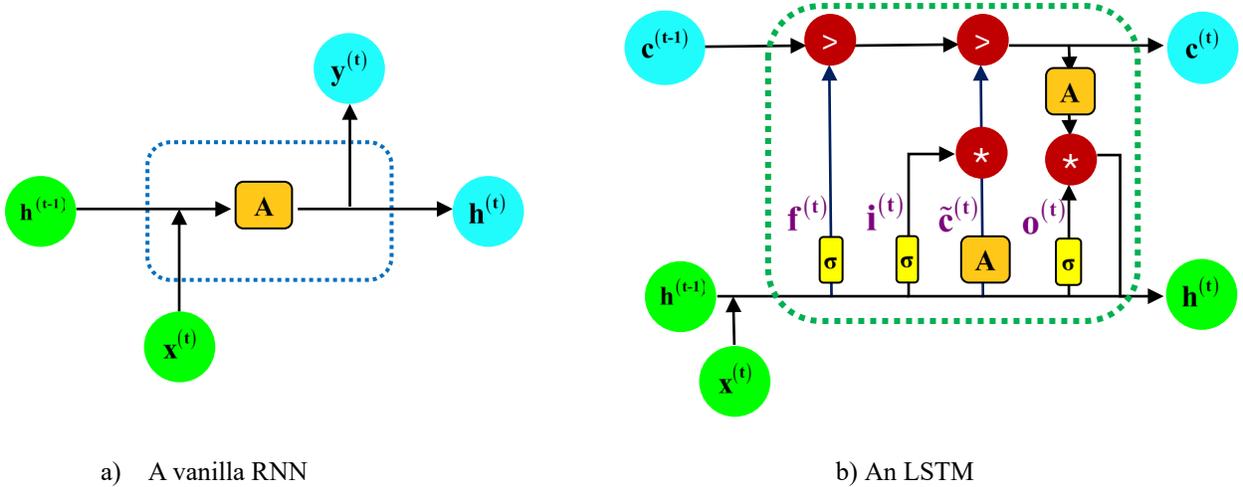

a) A vanilla RNN  b) An LSTM

Figure 3: Architecture of RNN

3.5.2 Incorporation of the second law of thermodynamics constraint in the RNN loss function

In our path-dependent material modeling approach based on RNNs, we enforce the Clausius-Duhem inequality (**Eq. 5a**), a key thermodynamics principle representing a fundamental expression of the second law, as a constraint for ensuring thermodynamic consistency. By incorporating this inequality, we impose a limitation on the evolution of internal variables and dissipation, ensuring that the predicted material behavior adheres to the second law of thermodynamics during isothermal processes. This approach enables the RNN-based model to maintain another physical principle besides frame invariance while capturing the complex, path-dependent nature of the material response. Following Holzapfel et al. [16], the dissipation inequality in **Eq. 5$_a$** for the internal state variable-based visco-hyperelastic framework can be expressed as

$$D_{int} = \sum \mathbf{Q}_\alpha : \dot{\mathbf{\Gamma}}_\alpha / 2 \geq 0 \tag{28}$$

Motivated by the Holzapfel [120] we use a specific form of the internal deviatoric history variables:

$$\mathbf{\Gamma}_\alpha = \frac{1}{\mu_\alpha} J^{-2/3} \text{Dev}\left(\nabla_{\bar{\mathbf{C}}} \Psi^\alpha(\bar{\mathbf{C}})\right) - \frac{1}{2\mu_\alpha} \mathbf{Q}_\alpha \tag{29}$$

where $\mu_\alpha$ denotes a non-negative, temperature-dependent parameter that is set to 1 in this study. While the data-driven model does not assume a specific constitutive law, we incorporate this form based on the Holzapfel Model [16] to satisfy the second law of thermodynamics. The expression balances stored energy and dissipation, with the first term capturing the isochoric stress response and the second term enforcing thermodynamic consistency through dissipation. Considering $\mathbf{S}_{iso}^\infty = J^{-2/3} \text{Dev}(2\nabla\Psi^\infty(\bar{\mathbf{C}}))$, and specific forms of $\mathbf{S}_{iso,\alpha} = \beta_\alpha^\infty \mathbf{S}_{iso}^\infty$ (in which $\beta_\alpha^\infty$ is a energy function) proposed by the Holzapfel Model [16] as well as the derivative of $\mathbf{\Gamma}_\alpha$, we get

$$\dot{\mathbf{\Gamma}}_\alpha = \frac{1}{2}\left(\beta_\alpha^\infty \dot{\mathbf{S}}_{iso}^\infty - \dot{\mathbf{Q}}_\alpha\right) \tag{30}$$

In a numerical setting, $\dot{\mathbf{\Gamma}}_\alpha$ at a given timestep (say, $t_{n+1}$) can be calculated via the finite difference method as

$$\dot{\mathbf{\Gamma}}_\alpha = \frac{1}{2\Delta t}\left[\beta_\alpha^\infty \left(\mathbf{S}_{iso}^\infty\big|_{t_{n+1}} - \mathbf{S}_{iso}^\infty\big|_{t_n}\right) - \left(\mathbf{Q}_\alpha\big|_{t_{n+1}} - \mathbf{Q}_\alpha\big|_{t_n}\right)\right] \tag{31}$$

Using the equilibrium isochoric stress prediction from the trained GPR model in **Eq. 31** and substituting the resulting equation into **Eq. 28** yields the viscous dissipation at a given timestep and deformation; the non-negativity of this viscous dissipation is incorporated into the loss function of our RNN model. **Figure 4** illustrates the workflow of the physics-informed loss function, which combines the mean squared error between predicted and true values with a dissipation penalty. The dissipation constraint ensures that the predicted dissipation remains non-negative; if this condition is violated, a penalty is added to the total loss, which is then used to update the RNN model weights during training.

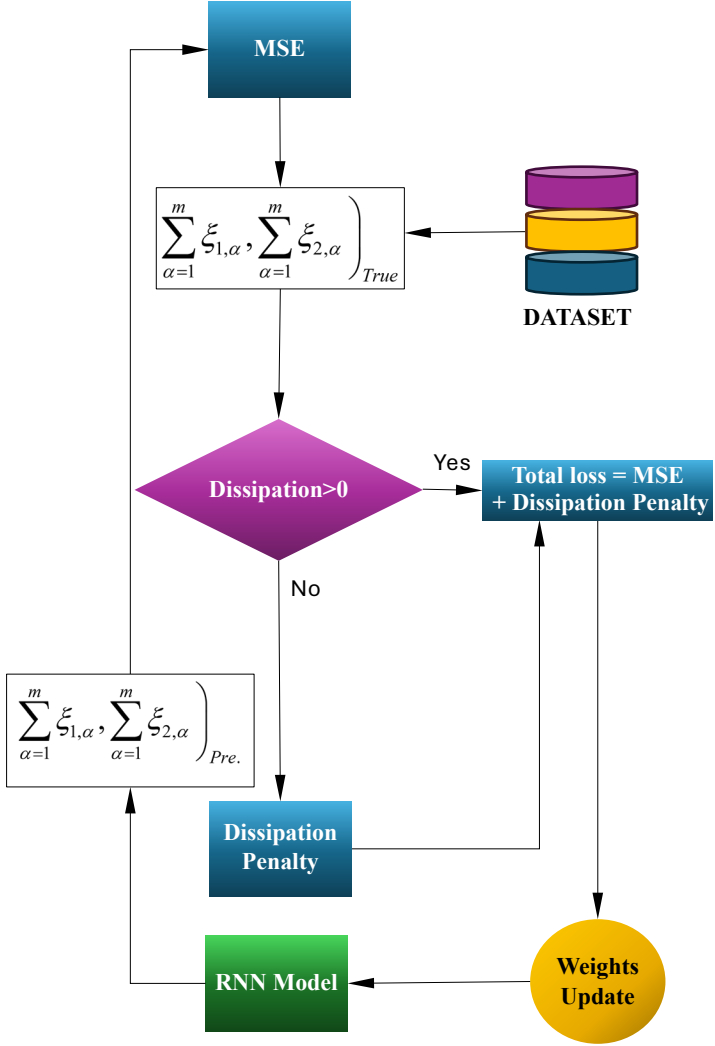

Figure 4: Flow diagrams of the RNN model with the combined physics-informed loss function

## 4. Model Evaluation and Performance

### 4.1 Training Dataset Generation

Our ML-based framework consists of three specialized surrogate models, employing a constrained RNN for viscoelastic behavior, and GPR models for volumetric and isochoric responses. Unlike previous data-driven models that treat different deformation modes as independent [106]our dataset is structured to maintain the inherent coupling between volumetric, isochoric, and viscoelastic responses, ensuring the integrity and consistency of the learned constitutive behavior. We use available specific constitutive models for the free and configurational energies to generate training data.

*4.1.1 Equilibrium part*

We investigate the material response under multiaxial loading conditions, where the principal stretch $\lambda_1$ differs from the other two, which are equal $\lambda_2=\lambda_3$. This deformation mode effectively captures 3D stress evolution and incorporates compressibility effects, which are critical in applications relevant to polymeric foams, soft elastomers, and polymers subjected to large hydrostatic pressure. Neo-Hookean energy function: $\Psi_{VOL}^{\infty}(J) = K/2(J-1)^2$, where $K=10$ is the bulk modulus, is used to generate the equilibrium volumetric stress component, while the Mooney-Rivlin strain energy function, $\Psi_{ISO}^{\infty}(\overline{\mathbf{C}}) = C_{10}(\overline{I}_1 - 3) + C_{01}(\overline{I}_2 - 3)$, with material parameters $C_{10}=10$ and $C_{01}=5$, is used to obtain the equilibrium isochoric stress component.

*4.1.2 Non-Equilibrium part*

For modeling nonlinear viscoelastic behavior under high stress or large deformations, the Holzapfel Model provides a robust framework based on differential equations that effectively capture nonlinear behavior and viscous dissipation. Thus, in the viscoelastic branches $\alpha$, $\mathbf{Q}_\alpha$ as the deviatoric internal variables correspond to the non-equilibrium stresses linked to $\Gamma_\alpha$ which is the internal deviatoric history variables. The following set of dissipative evolution equations are chosen to generate the training data of the non-equilibrium stress component:

$$\frac{d}{dt}(\mathbf{Q}_\alpha) + \frac{\mathbf{Q}_\alpha}{\tau_\alpha} = \frac{d}{dt}(\mathbf{S}_{ISO,\alpha}), \quad \alpha = 1,...,m \tag{25}$$

It is worth mentioning that the relaxational behavior of multiple chains can also be modeled using $m$ irreversible terms, each associated with a specific relaxation time $\tau_\alpha \in (0,\infty)$. Since polymers display viscoelastic behavior due to a medium composed of identical polymer chains, the isochoric part of the strain energy function $\beta_\infty^\alpha \in (0,\infty)$ can be derived from the volumetric part by incorporating the energy factor, thus, it is assumed that $\Psi_{ISO}^\alpha(\overline{\mathbf{C}}) = \beta_\infty^{\alpha\ 0} \Psi_{ISO}^{\infty}(\overline{\mathbf{C}})$.

$\mathbf{Q}_\alpha = \exp(-s/\tau_\alpha)\mathbf{Q}_\alpha^{0+} + \int_{0^+}^{t=s} \exp[-(s-t)/\tau_\alpha]\dot{\mathbf{S}}_{ISO,\alpha}\ dt$, is an explicit solution of **Eq. 25**, where $\mathbf{Q}_\alpha\big|_{t=0} = 0$ at initial time $t=0$. However, it needs refinement within the numerical analysis framework to implement a robust incremental algorithm for systematically updating the associated quantities. Therefore, considering $t_{n+1} = \Delta t + t_n$, we can obtain **Eq. 4** as

$$\mathbf{S} = J_{n+1} \frac{\partial \Psi_{\mathrm{VOL}}^{\infty}(J_{n+1})}{\partial J_{n+1}} \mathbf{C}_{n+1}^{-1} + 2 \frac{\partial^{0} \Psi_{\mathrm{ISO}}^{\infty}(\overline{\mathbf{C}}_{n+1}^{-1})}{\partial \mathbf{C}_{n+1}} + \sum_{\alpha=1}^{m} \mathbf{Q}_{\alpha,n+1} \tag{26}$$

$$\text{where } \mathbf{Q}_{\alpha,n+1} = \int_{0^{+}}^{t=s_{n}} \exp\left[-(s_{n}-t)/\tau_{\alpha}\right] \dot{\mathbf{S}}_{\mathrm{ISO},\alpha} \, dt + \int_{t=s_{n}}^{t=s_{n+1}} \exp\left[-(s_{n+1}-t)/\tau_{\alpha}\right] \dot{\mathbf{S}}_{\mathrm{ISO},\alpha} \, dt$$

Imposing the Mean Value Theorem on the split convolution form, it yields:

$$\mathbf{Q}_{\alpha,n+1} = \exp\left(-\frac{\Delta t}{\tau_{\alpha}}\right) \mathbf{Q}_{\alpha,n} + \exp\left(-\frac{\Delta t}{2\tau_{\alpha}}\right) \beta_{\infty}^{\alpha} \left(\mathbf{S}_{\mathrm{ISO},n+1}^{\infty} - \mathbf{S}_{\mathrm{ISO},n}^{\infty}\right) \tag{27}$$

To evaluate the model's performance across different time scales, we construct two datasets: one capturing short-term behavior and another capturing long-term response. These datasets, generated using different relaxation time parameters and numbers of Maxwell branches, reflect the typical range of viscoelastic behavior in soft materials to represent materials with varying relaxation characteristics.

- **Short-Term (ST) response:** The first dataset represents a material with four Maxwell branches, characterized by a relaxation time spectrum and uniform weighting factors. This configuration accounts for both fast and intermediate relaxation processes, allowing the model to describe complex, multi-phase viscoelastic responses. This model is well-suited for applications where both short-term energy dissipation and moderate-term stress relaxation are critical, such as biological tissues exposed to varying loading rates.
- **Long-Term (LT) response:** In contrast, the second dataset describes a material with a single Maxwell branch, defined by a long relaxation time. Due to subtle variations in stress response, it is an effective benchmark for evaluating the robustness and sensitivity of the surrogate model in capturing long-term, history-dependent behavior.

Incorporating these two datasets provides a comprehensive representation and a full range of viscoelastic behavior, encompassing both rapid and delayed stress relaxation processes through varying proportions of equilibrium and non-equilibrium stress contributions.

**4.2 GPR and LSTM-Based RNN Architecture Design**

In the ML analysis of the volumetric and isochoric components, the GPR model employs a kernel function composed of a constant kernel combined with a radial basis function kernel. This combination enables the model to capture both linear and nonlinear dependencies within the data. To enhance performance, the kernel hyperparameters are optimized through ten optimizer restarts, ensuring a more robust convergence. Additionally, a small regularization term ($\alpha = 10^{-4}$) is incorporated to improve numerical stability.

For the viscoelastic component, the RNN architecture consists of three LSTM layers with 128, 64, and 32 units, respectively, each utilizing the ReLU activation function. The model is trained using the Adam optimizer with a learning rate of 0.001, over 500 epochs. The batch size is set to 32 for the ST dataset and 64 for the LT dataset to ensure stable and efficient convergence. Both the input features and target values are normalized using min-max scaling and each training sample includes 5 (ST) or 6 (LT) consecutive time steps, allowing the LSTM layers to effectively model the time evolution of internal variables and capture the complex temporal behavior of the viscoelastic response.

### 4.3 Model Performance

We examine the material behavior under multiaxial one-cyclic loading, where the first principal stretch $\lambda_1$ differs from the other two, which are equal ($\lambda_2=\lambda_3$). To evaluate the model's generalizability and physical consistency, five cases are conducted as follows:

- Bulk and isochoric deformation modes with the GPR Model
- Different loading and unloading speeds for ST and LT datasets
- Different stretch levels for all three directions in tension and compression
- Adherence of the ML model to the second law of thermodynamics
- Noise sensitivity evaluation

To evaluate how well the surrogate models fit the data in each of the above cases, we use the percent relative error metric through $\left( \left\| \left( vec(\mathbf{S}) - vec(\tilde{\mathbf{S}}) \right) \right\| / \left\| vec(\tilde{\mathbf{S}}) \right\|_F \right) \times 100$, where $\|\cdot\|_F$ represents the Frobenius norm, $\tilde{\mathbf{S}}$ is the predicted stress tensor from the surrogate model, and $\mathbf{S}$ represents the true value of the stress tensor. The error is calculated at each data point, and the overall mean error across all data points is calculated as

$$ERR = \frac{\sum_{i=1}^{N} ERR_i}{N} \tag{24}$$

$N$ denotes the total number of data points in either the training or testing set.

#### 4.3.1 Case 1: Bulk and isochoric deformation modes with the GPR model

**Figures 5** and **6** show the longitudinal and transverse stress responses for the ST and LT datasets, with explicit decomposition into volumetric and isochoric parts using the GPR model. The ST dataset is generated under principal stretch values of $\lambda_1=1.28$ and $\lambda_2=\lambda_3=0.85$ for training, and $\lambda_1=1.5$ and $\lambda_2=\lambda_3=0.75$ for testing. Upon unloading, the stretches return to their original state. Moreover, for the LT dataset, the principal stretches are $\lambda_1=1.3$ and $\lambda_2=\lambda_3=0.75$, and for testing, they are $\lambda_1=1.5$ and $\lambda_2=\lambda_3=0.6$. While unloading, the stretches return to their original state. In both cases, the volumetric stress remains consistent, both in the training data and the test data beyond the training range. This indicates that the model captures the bulk (compressibility) response well. The equilibrium isochoric stress, which represents distortional behavior, increases more noticeably with stretching, especially in the test data beyond the training range.

For both ST and LT datasets, the model predictions match the training data closely and continue to give accurate results beyond the training range.

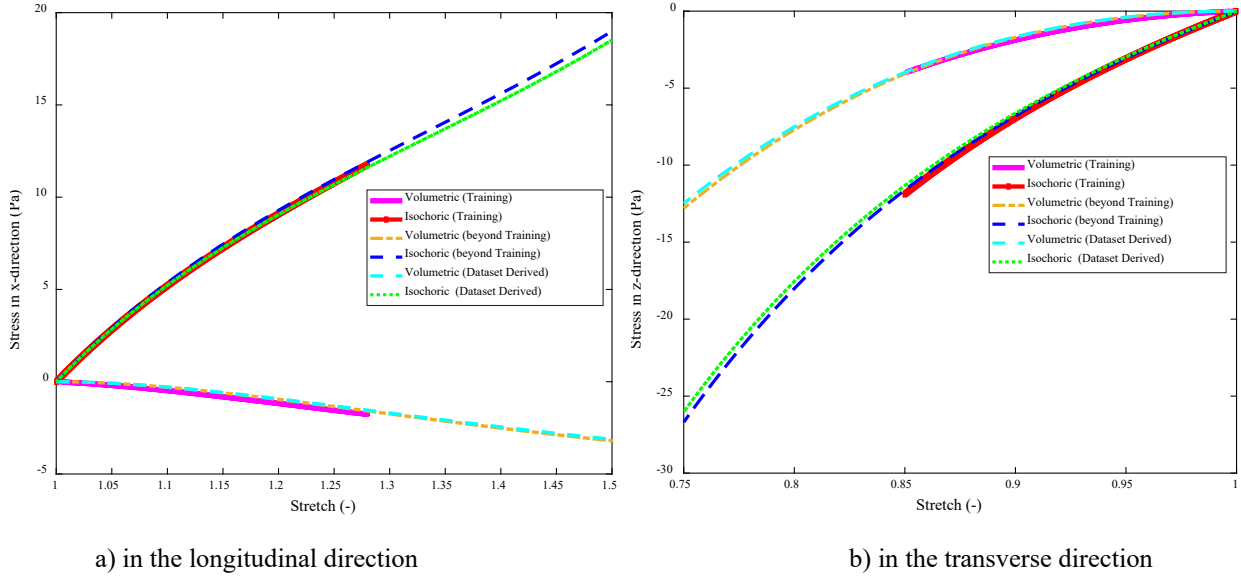

a) in the longitudinal direction

b) in the transverse direction

Figure 5: Evolution of the equilibrium volumetric and isochoric stress components in the ST dataset: Ground truth versus predicted responses.

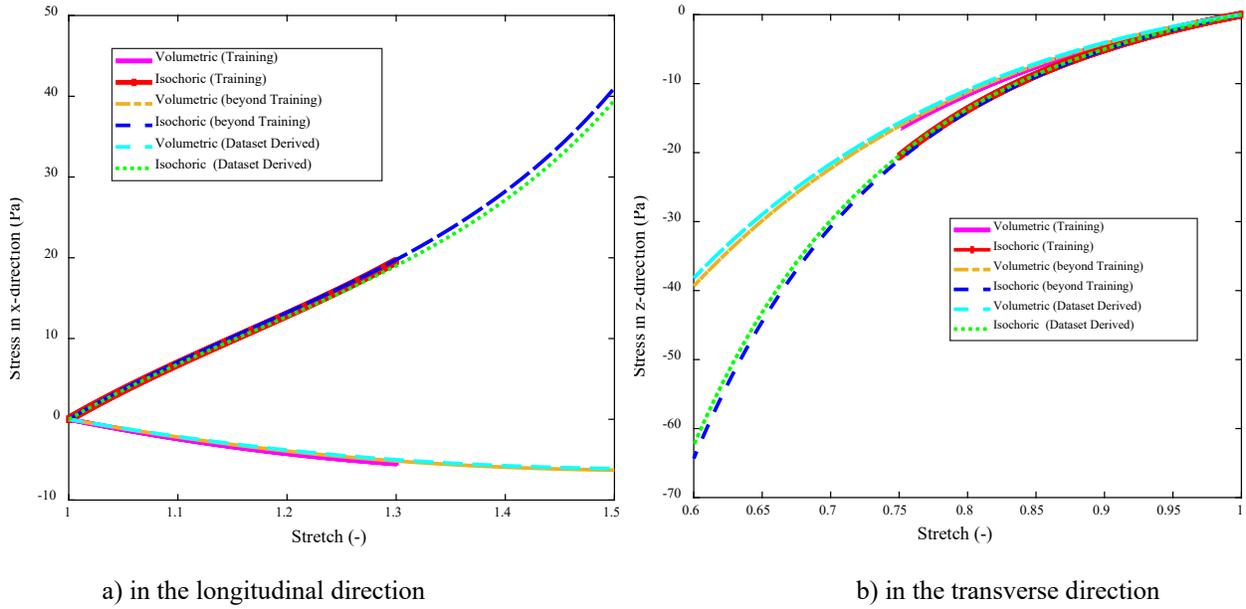

a) in the longitudinal direction

b) in the transverse direction

Figure 6: Evolution of the equilibrium volumetric and isochoric stress components in the LT dataset: Ground truth versus predicted responses.

Moreover, **Figures 7a** and **7b** illustrate the evolution of total stress in the x-direction over time for both the ST and LT datasets, considering different total time durations. In the ST case, the stress response shows a high initial peak, followed by a significant reduction in stress over time, indicating a strong viscoelastic relaxation effect. This remarkable decrease highlights the dominance of stress relaxation as the material rapidly dissipates stress in shorter total times. In contrast, the LT case exhibits a more gradual evolution of stress. While stress relaxation still occurs, the reduction in peak stress is not as significant as in ST, reflecting a slower viscoelastic response over longer durations. The dotted red curves in both figures represent asymptotic trends to the peak stresses, clearly showing the material response getting closer to quasi-static or equilibrium conditions. This trend emphasizes how, with increased total time, viscoelastic stress diminishes and reaches a steady-state response.

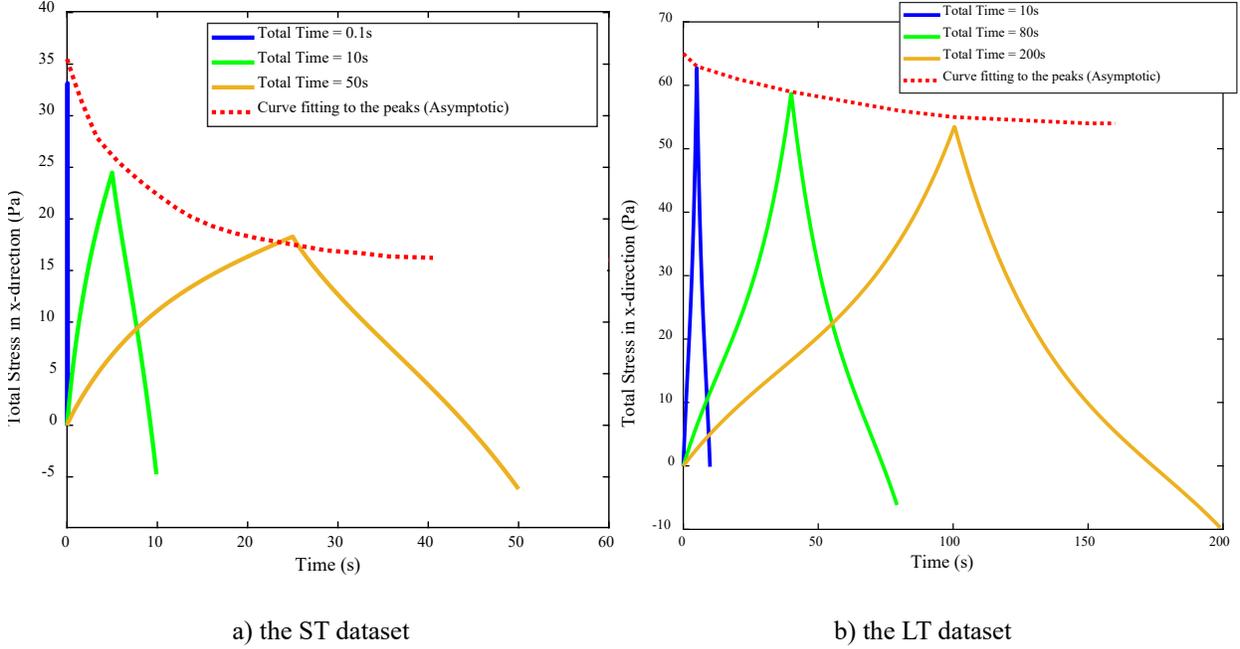

a) the ST dataset　　　　　　　　　　　　　　　　b) the LT dataset

Figure 7: Total stress-time behavior during stress relaxation to find the asymptotic line in the longitudinal direction

**4.3.2 Case2: Different loading and unloading rates for ST and LT datasets in $M_{visco}$ surrogate model**

In this section, we apply identical stretch levels of $\lambda_1=1.28$ and $\lambda_2=\lambda_3=0.85$ for both the training and testing phases. However, the total duration time will be varied from 1s to 10s for training and 0.1s to 25s for testing, leading to different loading and unloading rates. Moreover, for better clarity, we define the first term in **Eq. 14** ( $\text{vec}(\mathbb{G}_2)\sum \xi_{1,\alpha}(t,\overline{I}_1,\overline{I}_2)$ ) as the dominant instantaneous response, since it is based on the identity tensor (projected in a configuration-dependent way) and scaled by history-dependent functions. The second term ( $\text{vec}(\mathbb{G}_3)\sum \xi_{2,\alpha}(t,\overline{I}_1,\overline{I}_2)$ ) is called the dominant memory response, as it includes both time-dependent functions and the evolving internal configuration $\overline{\mathbf{C}}$, which captures the effect of past deformations. Although both terms are history-dependent, this terminology emphasizes the structural difference between a projected isotropic reference and a fully evolving memory tensor.

**Figures 8** and **9** illustrate the overall behavior of the ST dataset during the loading and unloading phases over time, shown for two different directions: x and z. During loading, both the immediate and memory responses increase, reflecting the accumulation of isochoric viscoelastic stress in the system. In the unloading phase, these responses decrease as the system relaxes. Throughout the entire process, the immediate response shows rapid changes, while the memory response evolves more gradually, capturing the system's time-dependent effects. To predict viscoelastic stress responses in x- and z- (or y-) directions, the ML model is trained and then tested on unseen data with a total time of 0.1s, 10s, and 25s (**Figure 8a-c**). The test durations extended over 90% beyond the training range in **Figure 8d-f**, from 0.1s (90% shorter than the shortest training duration) to 25s (more than 100% longer than the longest training range). Moreover, **Figure 9a–f** shows the training and testing of stress predictions in the z (or y) direction, capturing viscoelastic behavior across multiple stress components.

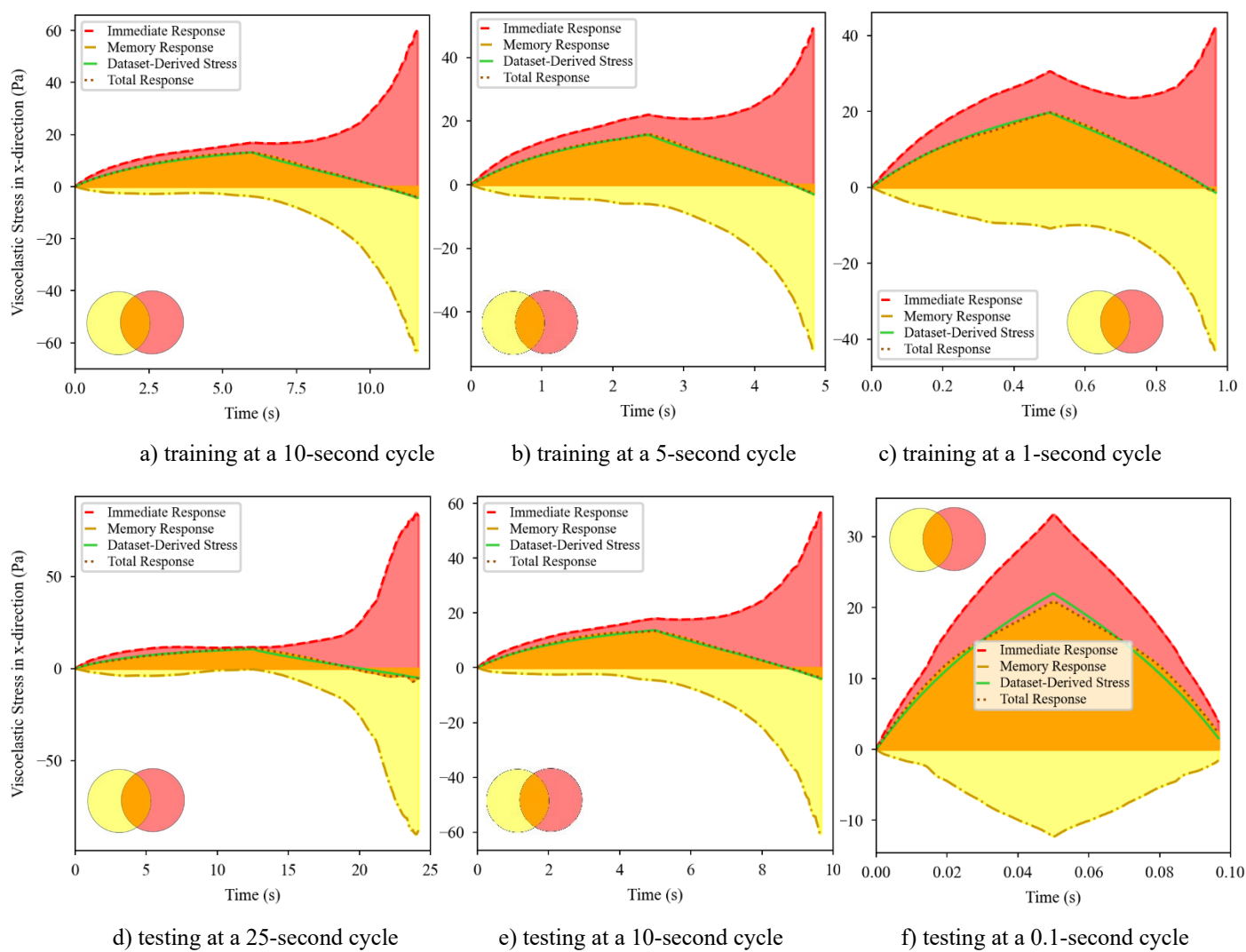

a) training at a 10-second cycle    b) training at a 5-second cycle    c) training at a 1-second cycle

d) testing at a 25-second cycle    e) testing at a 10-second cycle    f) testing at a 0.1-second cycle

Figure 8: Isochoric viscoelastic stress responses over time in the longitudinal (x) direction for the ST dataset

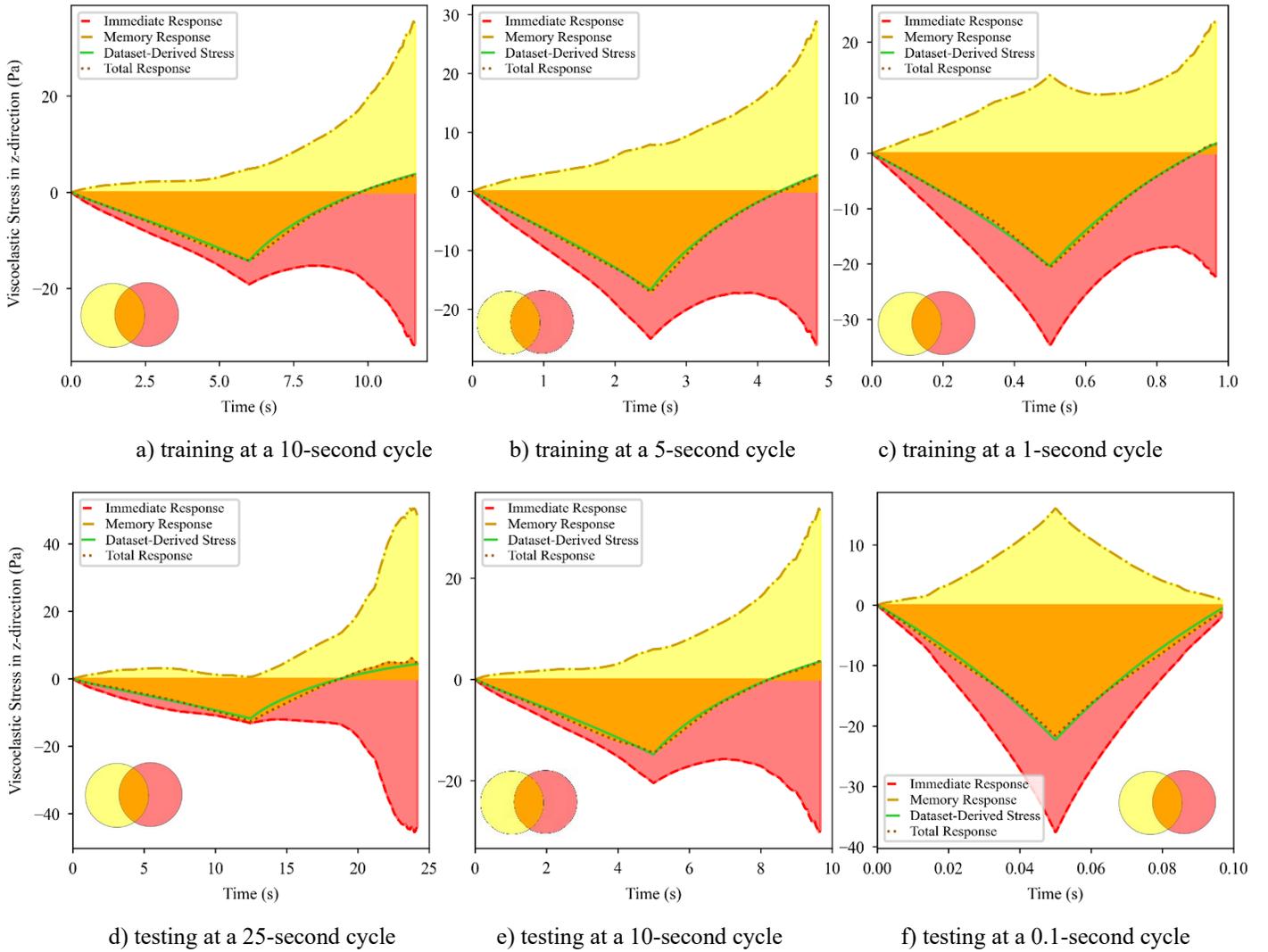

Figure 9: Isochoric viscoelastic stress responses over time in the transverse direction (z) for the ST dataset

From **Figures 8** and **9**, the ST dataset shows a quick drop in the isochoric viscoelastic stress component, mainly due to immediate effects and fast-fading memory. Meanwhile, the LT dataset, shown in **Figures 10** and **11**, shows that the memory response lasts longer over time. As a result, the total response decreases more slowly than in the ST dataset, indicating that the material relaxes more gradually due to longer-lasting memory effects. For the LT dataset, the surrogate ML viscoelastic model used the same stretch ranges during both training and testing (up to $\lambda_1=0.7$ and $\lambda_2=\lambda_3=1.3$). However, the total duration of the loading-unloading cycles differed: 30s, 60s, and 80s for training, and 10s, 100s, and 150s for testing. Despite significantly extending beyond the training range, the model accurately predicts isochoric viscoelastic stress responses in both the longitudinal and transverse directions, as depicted in the figures.

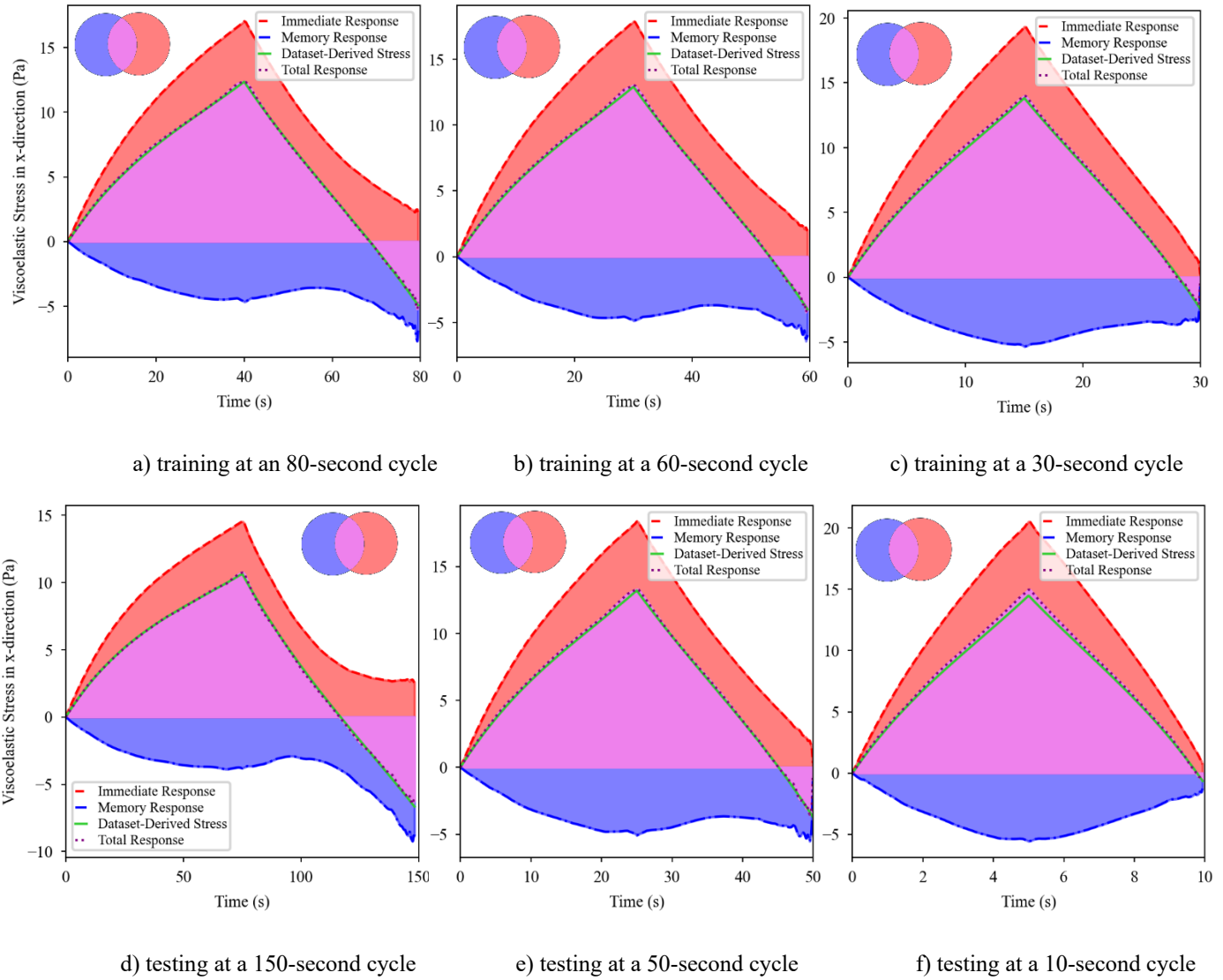

a) training at an 80-second cycle  b) training at a 60-second cycle  c) training at a 30-second cycle

d) testing at a 150-second cycle  e) testing at a 50-second cycle  f) testing at a 10-second cycle

Figure 10: Isochoric viscoelastic stress responses over time in the longitudinal (x) direction for the LT dataset

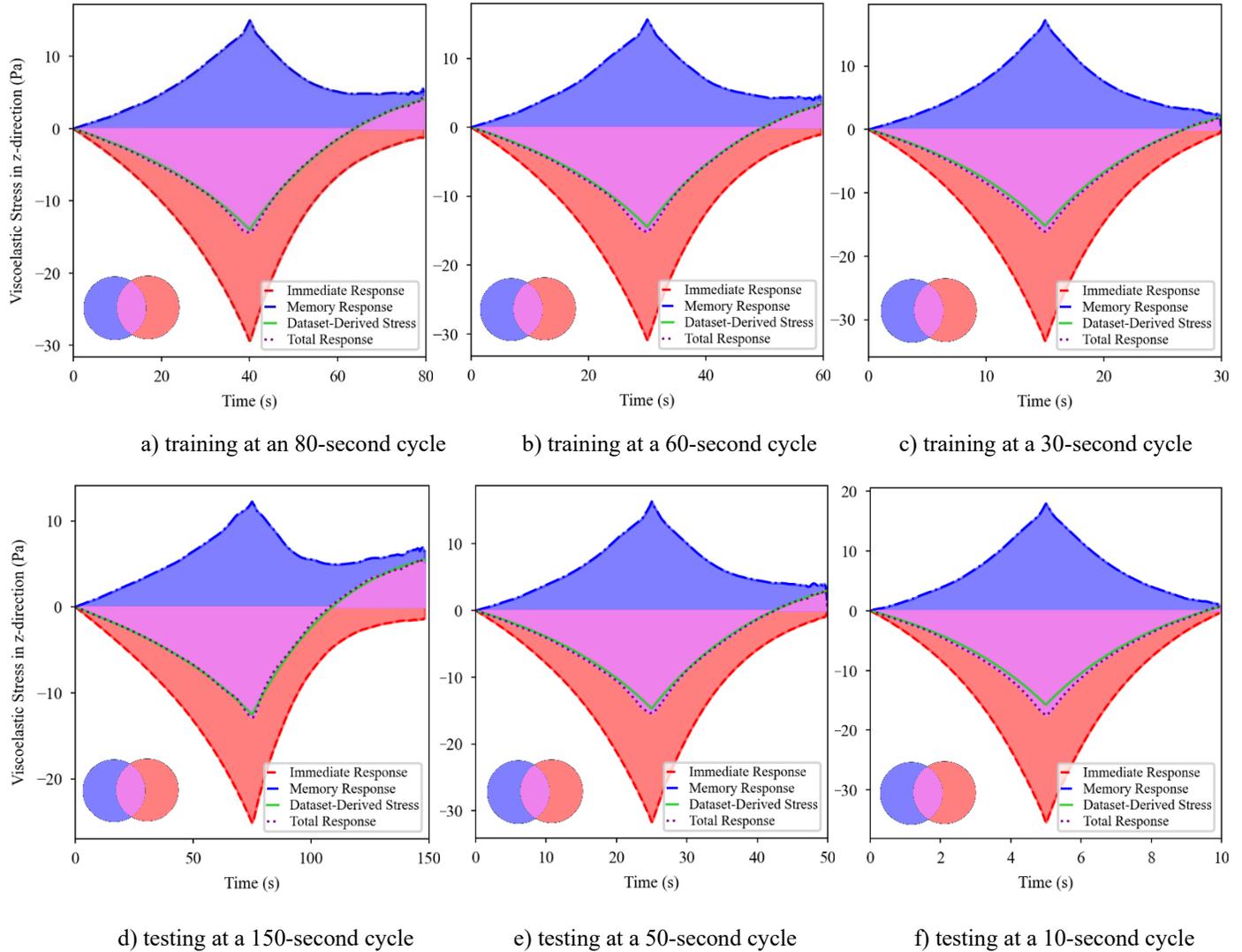

a) training at an 80-second cycle  b) training at a 60-second cycle  c) training at a 30-second cycle

d) testing at a 150-second cycle  e) testing at a 50-second cycle  f) testing at a 10-second cycle

Figure 11: Isochoric viscoelastic stress responses over time in the transverse (z) direction for LT dataset

**4.3.3 Case 3: Different stretch levels for all three directions in tension and compression**

In this case, the material is tested under variations of stretch levels to allow for a comprehensive investigation of how the viscoelastic material responds when it is subjected to a wide range of peak deformations and how the viscoelastic surrogate model performs at untrained stretch and strain rates simultaneously. Both the ST and LT datasets are used. **Figure 12** illustrates the predicted isochoric viscoelastic stress component in both the longitudinal (x) and transverse (z) directions and depicts contributions of the immediate and memory responses as well as the total response. The model was trained using principal stretch range of up to $\lambda_1=1.28$ and $\lambda_2=\lambda_3=0.85$, corresponding to moderate tension along the longitudinal axis and compression along the other two axes. The testing conditions include a scenario with $\lambda_1=0.8$ and $\lambda_2=\lambda_3=1.15$, representing a 37.5% reduction in the stretch along the primary axis and a 35.3% increase along the two transverse axes compared to the training scenario (**Figure 12a, b**). Another case involves $\lambda_1=0.7$ and $\lambda_2=\lambda_3=1.3$ in loading, representing a 45.3% decrease in $\lambda_1$ and a 52.9% increase in $\lambda_2=\lambda_3$, further pushing the model into extreme and highly non-proportional deformation conditions (**Figure 12c, d**). Finally, a scenario with $\lambda_1=1.5$ and $\lambda_2=\lambda_3=0.75$ applies substantial axial tension coupled with transverse compression, involving a 17.2% increase in the axial stretch and an 11.8% reduction in the transverse stretches (**Figure 12e, f**). For all the different conditions considered, the stretches return to their original undeformed state upon unloading.

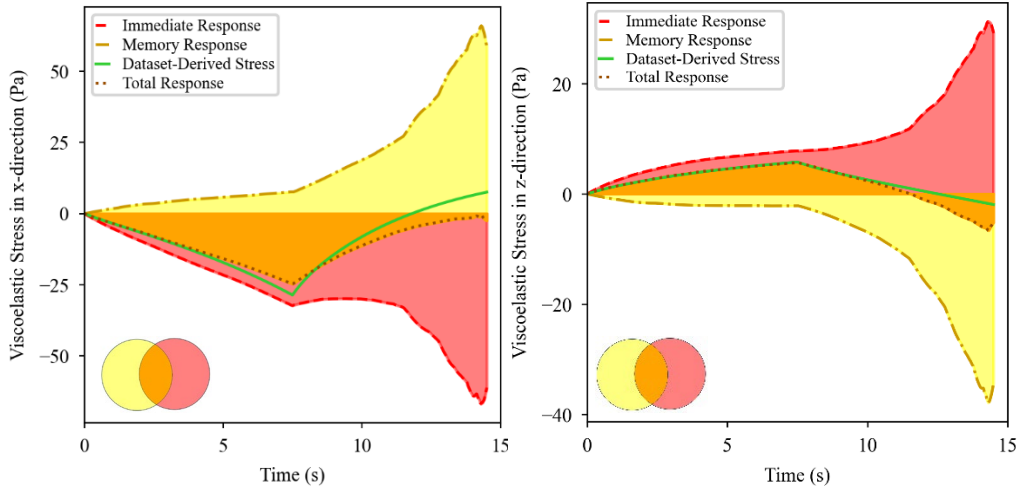

a) compression 0.8, x-direction, 15-second cycle    b) compression 1.15, z-direction, 15-second cycle

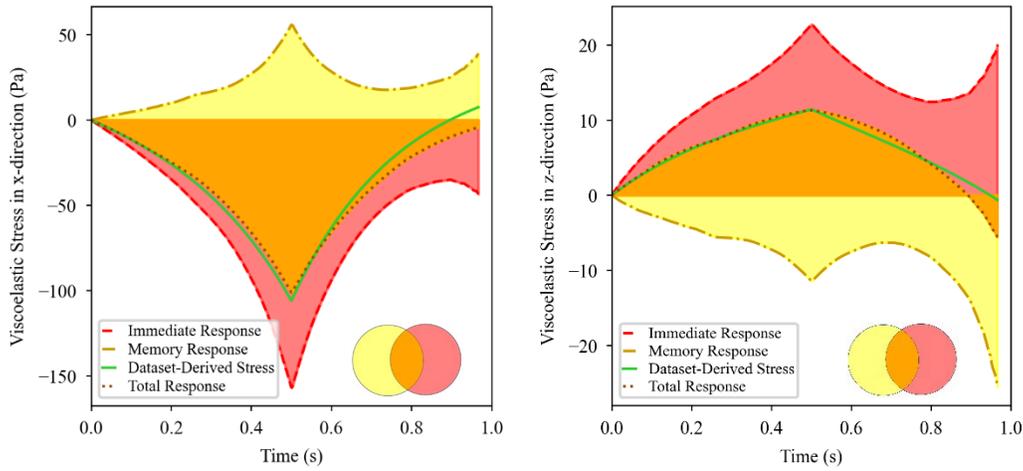

c) compression 0.7, x-direction, 10-second cycle    d) compression 1.3, z-direction, 10-second cycle

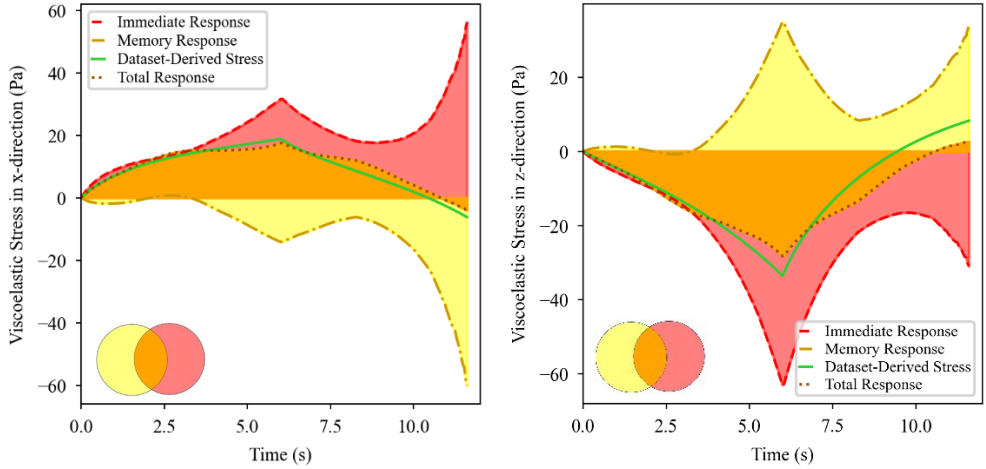

e) tension 1.5, x-direction, 10-second cycle     f) tension 0.75, z-direction, 10-second cycle

Figure 12: Isochoric viscoelastic stress responses in the x- and z-directions under various loading conditions for the ST dataset

**Figure 13** presents the predicted isochoric viscoelastic stress responses for the LT dataset in both the longitudinal (x) and the transverse (z) directions. This focuses on capturing the material's long-term viscoelastic relaxation behavior under sustained deformation. For the LT dataset, training is conducted under principal stretch values of $\lambda_1=1.3$ and $\lambda_2=\lambda_3=0.7$, representing a 3D loading state characterized by axial extension and transverse compression. The first test applied $\lambda_1=0.7$ and $\lambda_2=\lambda_3=1.3$, corresponding to a deformation state dominated by axial compression and transverse expansion. This represents a 46.2% decrease in axial stretch and an 85.7% increase in the transverse stretch compared to the training case. The total duration of this test is 10s, allowing for the evaluation of the material's rapid relaxation behavior following a sudden compressive load (**Figure 13a, b**). The second test imposes $\lambda_1=1.5$ and $\lambda_2=\lambda_3=0.6$, introducing a large axial tension combined with transverse compression (**Figure 13c, d**). This condition reflects a 15.4% increase in axial stretch and a 14.3% reduction in transverse stretch relative to the state of the LT training. The third test repeats the same deformation state but extends the total loading and unloading time to 180s (**Figure 13e,f**). This test evaluates long-term relaxation and the model's ability to predict stress under simultaneous untrained stretches and strain rates.

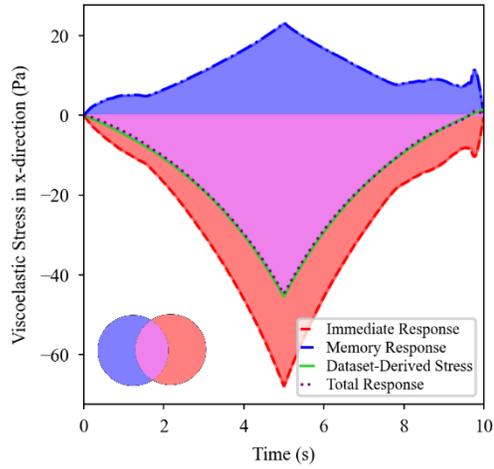
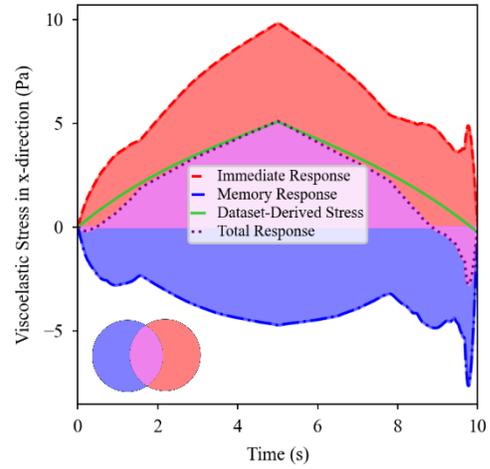

a) compression 0.7, x-direction, 10-second cycle   b) compression 1.3, z-direction, 10-second cycle

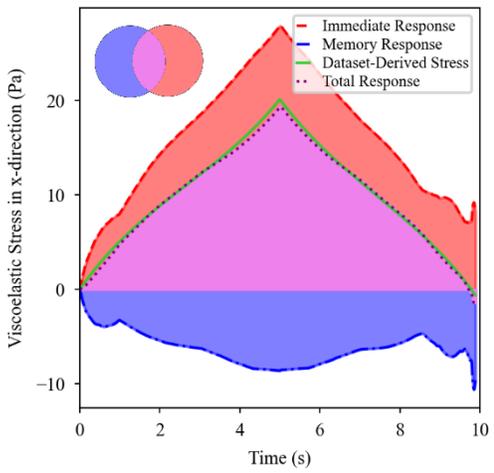
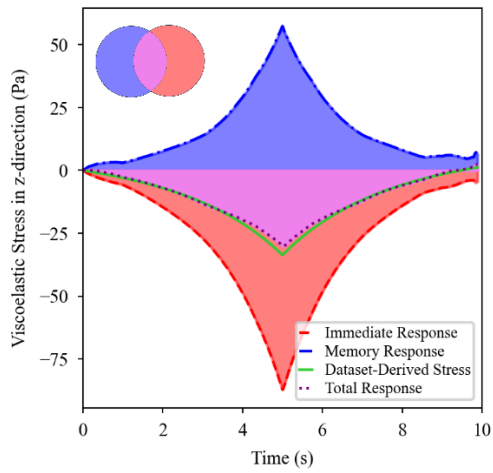

c) tension 1.5, x-direction, 10-second cycle   d) tension 0.6, z-direction, 10-second cycle

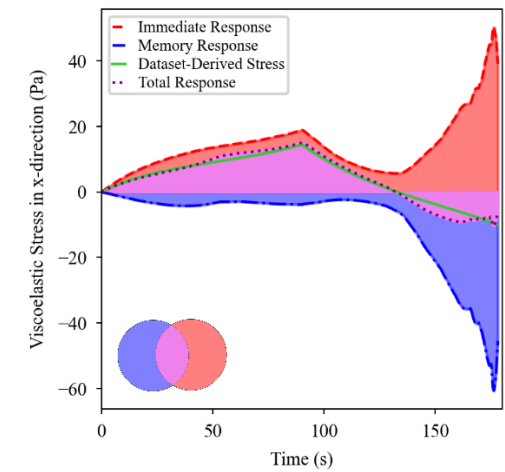
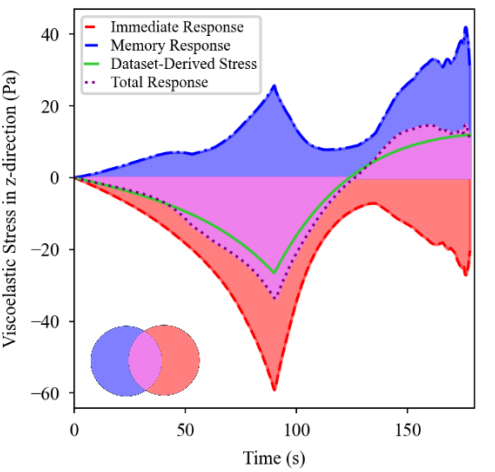

e) tension 1.5, x-direction, 180-second cycle   f) tension 0.6, z-direction, 180-second cycle

Figure 13: Isochoric viscoelastic stress responses in the x- and z-directions under various loading conditions for the LT dataset

The viscoelastic stress-stretch behavior in the longitudinal and transverse directions for both datasets is depicted in **Figure 14**. The ST dataset displays a more pronounced and immediate stress reaction to applied stretch during the 1-second cycle. In both tension and compression, the stress-stretch curves exhibit steeper slopes and more elastic-dominated behavior with minimal relaxation (i.e., smaller hysteresis). At a stretch ratio of 1.35 (tension), the x-direction stress reaches approximately 30 Pa, reflecting the material's high stiffness under tensile loading (**Figure 14a**). Conversely, in compression at a stretch ratio of 0.7, the stress plunges to around -110 Pa, indicating a rapid and significant resistance to compressive deformation. These behaviors are characteristic of materials dominated by immediate elastic response, where stress magnitudes are higher due to limited time for relaxation. However, the LT dataset demonstrates a more gradual stress-stretch response. Compared to the ST case, the curves are less steep, reflecting the time-dependent, viscoelastic relaxation effects during the 50-second cycle (**Figure 14b**). This energy dissipation over time results in a more compliant behavior, with stress responses that are more moderate and less sensitive to instantaneous deformation.

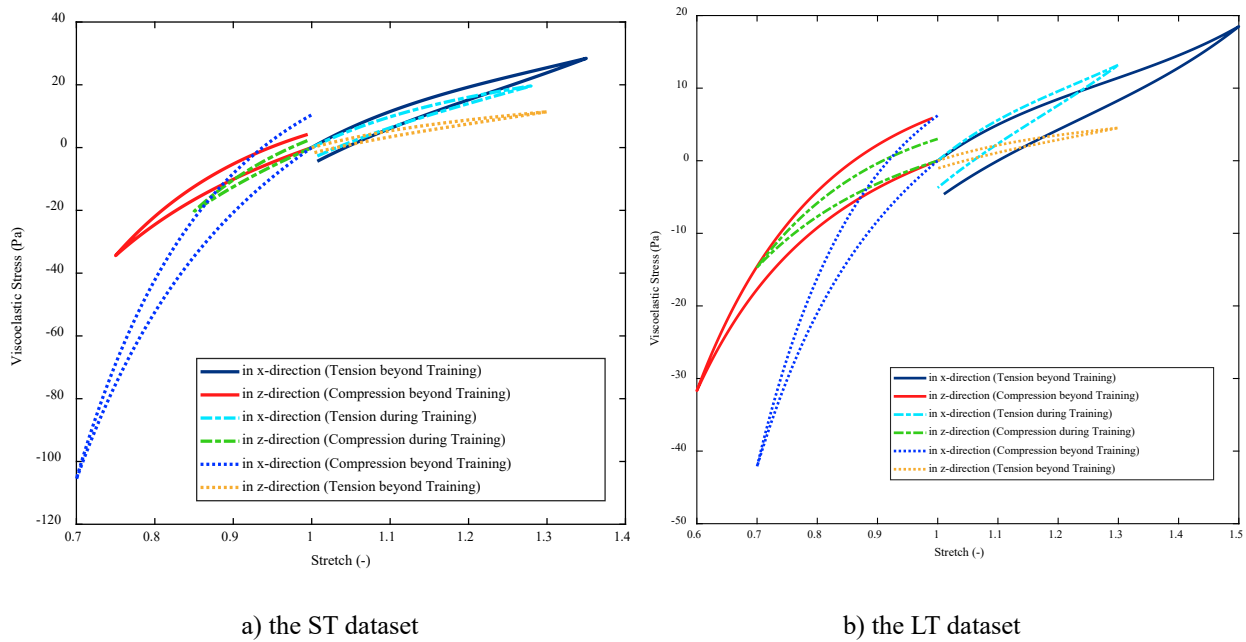

a) the ST dataset  b) the LT dataset

Figure 14: Isochoric viscoelastic tress-stretch behavior in the x- and z-directions subjected to tension and compression

### 4.3.4 Case 4: Adherence of the ML model to the second law of thermodynamics

The second law of thermodynamics requires that viscous dissipation in a material be non-negative(i.e., the Clausius-Duhem inequality). In the context of viscoelastic modeling, this condition is typically reflected through the dissipation rate (calculated using **Eq. 28**) remaining non-negative throughout the loading-unloading cycles. For the ST dataset, the dissipation rates shown in **Figure 15a**, display sharp, high-intensity peaks at the early stages of the loading cycles. The dissipation magnitude is highest for the shortest total time of the 1-second cycle, reaching up to 300 W/m³, and decreases progressively as the total time increases to 30s. The dissipation rates remain strictly non-negative throughout the process, satisfying the thermodynamic requirement. The pronounced peaks at shorter times indicate a rapid viscous dissipation due to the material's quick relaxation behavior. Additionally, for longer durations within the ST dataset, the dissipation is more moderate and stabilizes at lower values, confirming the model's thermodynamic consistency over a range of loading times. In the LT dataset, the dissipation rates exhibit a different profile, as shown in **Figure 15b**. Here, the dissipation peaks are significantly lower in magnitude compared to the ST dataset, with the highest dissipation reaching approximately 25 W/m³ for the 1-second cycle. As the total time increases to 180s, the dissipation profiles show a gradual rise and fall over extended periods, characteristic of a material with long-term memory effects and slower energy dissipation. The spread of dissipation over longer time scales reflects a gradual and smaller entropy production and slower relaxation behavior inherent in the LT dataset.

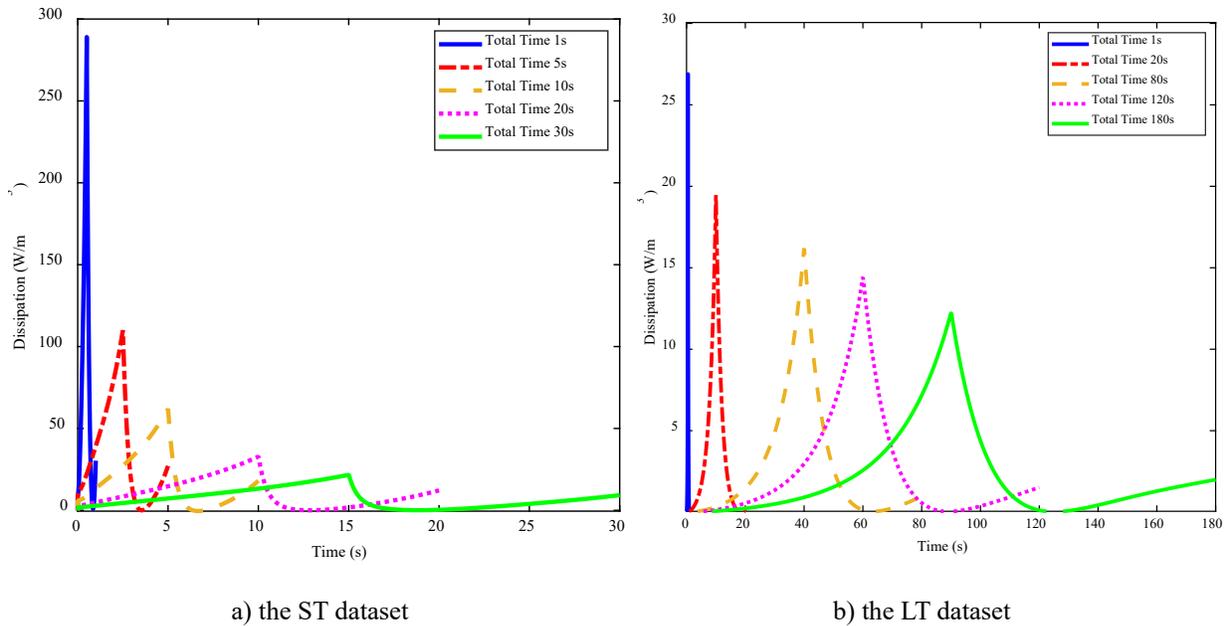

a) the ST dataset　　　　　　　　　　　　　　b) the LT dataset

Figure 15: Dissipation-time behavior for both datasets at different total times

### 4.3.5 Case 5: Noise Sensitivity Evaluation

The uncertainty handling capability of the RNN-physic-based ML model is assessed by introducing synthetic experimental noise at a level of approximately 6%. For both the ST and LT datasets, the viscoelastic surrogate model is trained and evaluated on both noise-free and noisy data. **Figures 16** and **17** illustrate that the model accurately predicted viscoelastic stress responses in both the x- and z-directions, consistently capturing the loading and unloading behavior. Despite the presence of noise, the predictions remained in close agreement with both the synthetic data and the noise-free baseline responses. In the LT dataset, which involves more complex and delayed stress relaxation behavior, the model maintained reliable accuracy, with only minor fluctuations observed.

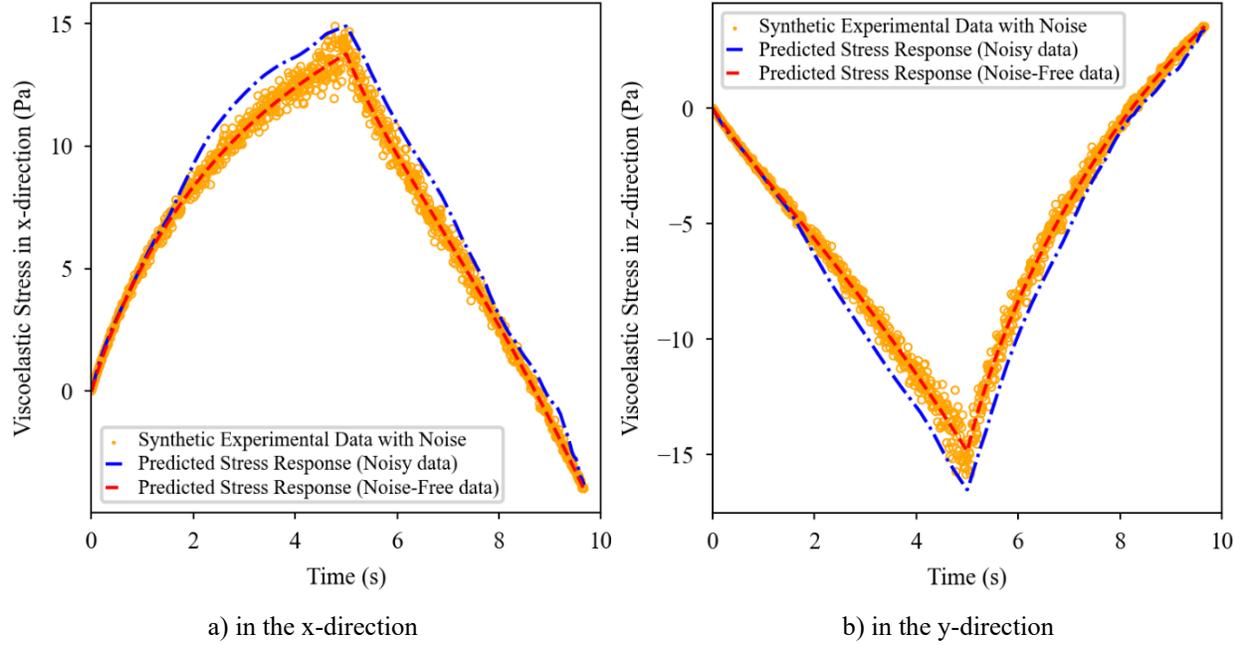

a) in the x-direction	b) in the y-direction

Figure 16: Comparison between synthetic experimental data with noise and noise-free responses for the ST dataset

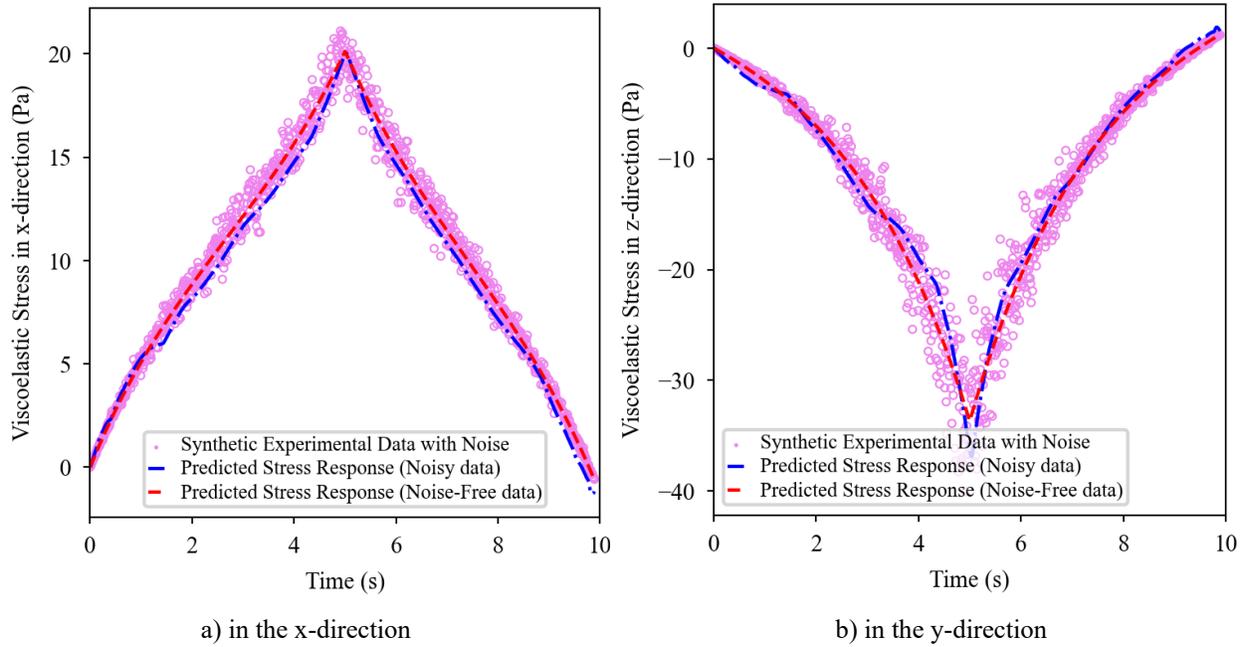

a) in the x-direction	b) in the y-direction

Figure 17: Comparison between synthetic experimental data with noise and noise-free responses for the LT dataset

## 5. Conclusions

This study presents a hybrid physics-informed machine learning framework for modeling nonlinear, visco-hyperelastic, history-dependent material behavior under multiaxial cyclic loading. Built upon a generalized internal state variable-based visco-hyperelastic formulation, the framework couples GPR for learning equilibrium responses with LSTM-based RNNs to capture time-dependent viscoelastic stress evolution. After developing the general surrogate model, we employed the neo-Hookean volumetric energy density, the Mooney-Rivlin strain energy density, and the nonlinear Holzapfel differential viscoelastic model to generate synthetic datasets representing both short-term and long-term relaxation behaviors. The model was rigorously tested across a wide range of stretch conditions. Training was performed using moderate deformation states (e.g., $\lambda_1 = 1.28$, $\lambda_2 = \lambda_3 = 0.85$), while testing included significantly untrained and extreme cases, such as $\lambda_1 = 1.5$ and $\lambda_2 = \lambda_3 = 0.75$ (axial tension and transverse compression), $\lambda_1 = 0.7$ and $\lambda_2 = \lambda_3 = 1.3$ (axial compression and transverse tension), and $\lambda_1 = 0.8$ and $\lambda_2 = \lambda_3 = 1.15$. The model showed excellent predictive capability across these multiaxial loading paths, accurately predicting volumetric, isochoric hyperelastic, and isochoric viscoelastic stress components as well as the total stress. For instance, under $\lambda_1 = 1.35$ (tension), the predicted viscoelastic stress reached 30 Pa in the x-direction for the ST case, while under $\lambda_1 = 0.7$ (compression), the stress dropped to –110 Pa, capturing sharp asymmetries between tensile and compressive responses. In the LT dataset, predictions were more compliant, with smoother stress-stretch curves under extended time durations (e.g., 180 s). Dissipation rates also remained strictly non-negative, confirming thermodynamic consistency, with peak values around 300 W/m³ for fast ST cycles and ~25 W/m³ for slow LT responses. Even under 6% synthetic noise, the model preserved accuracy across all stretch levels and directions. Altogether, this stretch-informed validation confirms that the proposed framework not only generalizes well beyond its training domain but also robustly captures complex, nonlinear, and history-dependent material behavior under diverse deformation paths.

**Acknowledgment**

The authors acknowledge the support of the US National Science Foundation under Grant Numbers OIA-1946231 and 2331294 as well as the Louisiana Board of Regents for the Louisiana Materials Design Alliance (LAMDA).